\documentclass[sort&compress]{elsarticle}
\usepackage[english]{babel}
\usepackage{graphicx,rotating,multirow}
\usepackage{bm}
\usepackage{amsmath,multirow,mathrsfs}
\usepackage{amssymb}
\usepackage{cuted}
\journal{Physics Letters A}

\newcommand{\AJP}{ Am. J. Phys. }
\newcommand{\AM}{ Annals Math. }

\newcommand{\APB}{ Ann. Phys. (Berlin) }
\newcommand{\APNY}{ Ann. Phys. (N.Y.) }

\newcommand{\CMP}{ Commun. Math. Phys. }
\newcommand{\CPB}{ Chin. Phys. B }

\newcommand{\CRA}{ C. R. Acad. Sci. Ser. A }
\newcommand{\EJP}{ Eur. J. Phys. }

\newcommand{\IJMPB}{ Int. J. Mod. Phys. B }

\newcommand{\JAP}{ J. Appl. Phys. }

\newcommand{\JSP}{ J. Stat. Phys. }

\newcommand{\PA}{ Physica A }

\newcommand{\PLA}{ Phys. Lett. A }
\newcommand{\PR}{ Phys. Rev. }
\newcommand{\PRA}{ Phys. Rev. A }
\newcommand{\PRB}{ Phys. Rev. B }

\newcommand{\PRE}{ Phys. Rev. E }
\newcommand{\PRL}{ Phys. Rev. Lett. }

\newcommand{\PS}{ Phys. Scr. }

\newcommand{\RMP}{ Rev. Mod. Phys. }

\newcommand{\SST}{ Semicond. Sci. Technol. }

\begin{document}
\begin{frontmatter}
\title{Quantum information measures of the Aharonov-Bohm ring in uniform magnetic fields}
\author{O. Olendski}
\ead{oolendski@sharjah.ac.ae}
\address{Department of Applied Physics and Astronomy, University of Sharjah, P.O. Box 27272, Sharjah, United Arab Emirates}
\begin{abstract}
Shannon quantum information entropies $S_{\rho,\gamma}$, Fisher informations $I_{\rho,\gamma}$, Onicescu energies $O_{\rho,\gamma}$ and complexities $e^SO$ are calculated both in the position (subscript $\rho$) and momentum ($\gamma$) spaces for the azimuthally symmetric two-dimensional nanoring that is placed into the combination of the transverse uniform magnetic field $\bf B$ and the Aharonov-Bohm (AB) flux $\phi_{AB}$ and whose  potential profile is modeled by the superposition of the quadratic and inverse quadratic dependencies on the radius $r$. The increasing intensity $B$ flattens momentum waveforms $\Phi_{nm}({\bf k})$ and in the limit of the infinitely large fields they turn to zero: $\Phi_{nm}({\bf k})\rightarrow0$ at $B\rightarrow\infty$, what means that the position wave functions $\Psi_{nm}({\bf r})$, which are their Fourier counterparts, tend in this limit to the $\delta$-functions. Position (momentum) Shannon entropy depends on the field $B$ as a negative (positive) logarithm of $\omega_{eff}\equiv\left(\omega_0^2+\omega_c^2/4\right)^{1/2}$, where  $\omega_0$ determines the quadratic steepness of the confining potential and $\omega_c$ is a cyclotron frequency. This makes the sum ${S_\rho}_{nm}+{S_\gamma}_{nm}$ a field-independent quantity that increases with the principal $n$ and azimuthal $m$ quantum numbers and does satisfy entropic uncertainty relation. Position Fisher information does not depend on $m$, linearly increases with $n$ and varies as $\omega_{eff}$ whereas its $n$ and $m$ dependent Onicescu counterpart ${O_\rho}_{nm}$ changes as $\omega_{eff}^{-1}$. The products ${I_\rho}_{nm}{I_\gamma}_{nm}$ and ${O_\rho}_{nm}{O_\gamma}_{nm}$ are $B$-independent quantities. A dependence of the measures on the ring geometry is discussed. It is argued that a variation of the position Shannon entropy or Onicescu energy with the AB field uniquely determines an associated persistent current as a function of $\phi_{AB}$ at $B=0$. An inverse statement is correct too.
\end{abstract}
\end{frontmatter}

\section{Introduction}\label{sec1}
Magnetic field influence on the properties of the quantum rings (QRs) presents one of the most interesting problems in the physics of these artificial structures with nontrivial topology \cite{Fomin1}. In addition to the traditional substances, such as superconductors \cite{Deaver1,Byers1,Doll1}, normal metals \cite{Levy1,Chandrasekhar1} and semiconductors \cite{Mailly1,Liu1,Fuhrer1,Keyser1,Giesbers1,Kleemans1}, from which they are made, relatively new materials, such as graphene, are  being actively used in the fabrication of the QRs too \cite{Russo1}. An extreme configuration of the nonuniform field that is transverse to the ring plane is a geometry when the whole magnetic flux is concentrated in the centre of the QR. Despite the fact that this Aharonov-Bohm (AB) intensity \cite{Aharonov1} is not accessible for the charged quantum particles moving inside the ring, it strongly affects their properties leading, in particular, to the QR persistent currents \cite{Buttiker1} flowing along the azimuthal direction of the structure.

Among different characteristics of the nanosystems, a very special role is played by quantum-information measures, which are descriptors of miscellaneous facets of the distribution of position $\rho({\bf r})$ and wave vector $\gamma({\bf k})$ densities that are squared magnitudes of the corresponding waveforms $\Psi_n({\bf r})$ and $\Phi_n({\bf k})$:
\begin{subequations}\label{DensityXK1}
\begin{align}\label{DensityX1}\rho_n({\bf r})&=\left|\Psi_n({\bf r})\right|^2\\
\label{DensityK1}\gamma_n({\bf k})&=\left|\Phi_n({\bf k})\right|^2,
\end{align}
\end{subequations} 
where nonnegative integer index $n$ counts all bound orbitals of the structure, and functions $\Psi_n$ and $\Phi_n$ are related through the Fourier transformations:
\begin{subequations}\label{Fourier1}
\begin{align}\label{Fourier1_1}
\Phi_n({\bf k})&=\frac{1}{(2\pi)^{l/2}}\int_{\mathbb{R}^l}\Psi_n({\bf r})e^{-i{\bf kr}}d{\bf r},\\
\label{Fourier1_2}
\Psi_n({\bf r})&=\frac{1}{(2\pi)^{l/2}}\int_{\mathbb{R}^l}\Phi_n({\bf k})e^{i{\bf rk}}d{\bf k},
\end{align} 
\end{subequations}
with the integration carried out over the whole available $l$-dimensional position, Eq.~\eqref{Fourier1_1}, or momentum, Eq.~\eqref{Fourier1_2}, region under the assumption that the waveforms are orthonormalized:
\begin{equation}\label{OrthoNormality1}
\int_{\mathbb{R}^l}\Psi_{n'}^*({\bf r})\Psi_n({\bf r})d{\bf r}=\int_{\mathbb{R}^l}\Phi_{n'}^\ast({\bf k})\Phi_n({\bf k})d{\bf k}=\delta_{nn'},
\end{equation}
where $\delta_{nn'}=\left\{\begin{array}{cc}
1,&n=n'\\
0,&n\neq n'
\end{array}\right.$ is a Kronecker delta, $n,n'=0,1,2,\ldots$.

Position $S_\rho$ and momentum $S_\gamma$ Shannon quantum-information entropies defined as
\begin{subequations}\label{Shannon1}
\begin{align}\label{Shannon1_R}
S_\rho&=-\int_{\mathbb{R}^l}\rho({\bf r})\ln\rho({\bf r})d{\bf r}\\
\label{Shannon1_P}
S_\gamma&=-\int_{\mathbb{R}^l}\gamma({\bf k})\ln\gamma({\bf k})d{\bf k},
\end{align}
\end{subequations}
describe quantitatively the lack of our knowledge about location and motion of the particle: the smaller (larger) either of them is, the more (less) information is available about the corresponding property. These two functionals are not independent from each other but are related as \cite{Bialynicki1,Beckner1}
\begin{equation}\label{EntropyInequality1}
S_\rho+S_\gamma\geq l(1+\ln\pi).
\end{equation}
A comparison of this inequality with the Heisenberg uncertainty relation reveals that Eq.~\eqref{EntropyInequality1} presents a much more general base for defining 'uncertainty' \cite{Coles1} since the former always holds true whereas the latter is either violated or meaningless, among others, for the structures with non-Dirichlet boundary conditions \cite{Bialynicki3,Olendski5,Olendski1,Olendski2}. The significance of the Shannon entropies stimulates their intensive detailed analysis for different geometries, such as one-dimensional (1D) harmonic oscillator \cite{Gadre1} and its multidimensional generalizations \cite{Toranzo1,Yanez1}, 3D \cite{Gadre1} and multidimensional hydrogen atom \cite{Yanez1}, miscellaneous 1D potentials: one or two Dirac $\delta$-functions \cite{Bouvrie1}, P\"{o}schl-Teller \cite{Sun3}, hyperbolic \cite{ValenciaTorres1}, squared tangent \cite{Dong1}, symmetric \cite{Sun1} and asymmetric \cite{Sun2} trigonometric Rosen-Morse potentials, systems with position-dependent mass \cite{YanezNavarro1}, and many, many others. The literature of the Shannon entropies, as well as other measures mentioned below, is growing at a very impressive rate.

Expressions for the position $I_\rho$ and momentum $I_\gamma$ Fisher informations \cite{Fisher1,Frieden1} read:
\begin{subequations}\label{Fisher1}
\begin{align}\label{Fisher1_R}
I_\rho&=\int_{\mathbb{R}^l}\frac{|{\bf\nabla}_{\bf r}\rho({\bf r})|^2}{\rho({\bf r})}d{\bf r}\\
\label{Fisher1_P}
I_\gamma&=\int_{\mathbb{R}^l}\frac{|{\bf\nabla}_{\bf k}\gamma({\bf k})|^2}{\gamma({\bf k})}d{\bf k}.
\end{align}
\end{subequations}
A presence of the gradients makes these functionals local measures of uncertainty that are sensitive to the speed of change of the corresponding density. Contrary to the sum of the Shannon entropies, the product of two Fisher informations does not have a rigorous lower bound: widely discussed inequality 
\begin{equation}\label{FisherInequality1}
I_\rho I_\gamma\geq4
\end{equation}
is violated, e.g., for the non-Dirichlet systems \cite{Olendski1,Olendski2}.

Physical meaning of the Onicescu energies \cite{Onicescu1}, or disequilibria,
\begin{subequations}\label{Onicescu1}
\begin{align}\label{Onicescu1_R}
O_\rho&=\int_{\mathbb{R}^l}\rho^2({\bf r})d{\bf r}\\
\label{Onicescu1_P}
O_\gamma&=\int_{\mathbb{R}^l}\gamma^2({\bf k})d{\bf k}
\end{align}
\end{subequations}
lies in the fact that they describe deviations from the uniform distributions.

Based on these functionals, some combined measures could be defined; for example, a complexity
\begin{equation}\label{CGL1}
CGL=e^SO
\end{equation}
introduced by R. G. Catal\'{a}n, J. Garay and R. L\'{o}pez-Ruiz \cite{Catalan1} tries to simultaneously quantify an uncertainty represented by the first factor as well as disequilibrium due to the presence of the Onicescu energy. Remarkably, for any $l$-dimensional space either position or momentum component of this dimensionless quantity can not be smaller than unity \cite{LopezRosa1}
\begin{equation}\label{CGLinequality1}
CGL\geq1.
\end{equation}

In this Letter, an exact analysis of the above mentioned quantum-information measures is provided for the QR in the superposition of the transverse uniform magnetic field $\bf B$ and the Aharonov-Bohm flux $\phi_{AB}$. For the azimuthally symmetric 2D potential profile $U({\bf r})$ of the ring [with ${\bf r}=\left(r,\varphi_r\right)$ being position polar coordinates with the origin of this frame of reference coinciding with the geometrical center of the structure] we use a combination of the quadratic and inverse quadratic depedendencies on the radius $r$ \cite{Bogachek1,Tan1,Tan2,Tan3,Fukuyama1,Bulaev1,Simonin1,Margulis1,Olendski3,Xiao1,Negrete1}:
\begin{equation}\label{Potential1}
U(r)=\frac{1}{2}m^*\omega_0^2r^2+\frac{\hbar^2}{2m^*r^2}\,a-\hbar\omega_0a^{1/2},
\end{equation}
where $m^*$ is an effective mass of a charge carrier, frequency $\omega_0$ defines a steepness of the confining in-plane outer surface of the QR with its characteristic radius $r_0=[\hbar/(2m^*\omega_0)]^{1/2}$, and positive dimensionless constant $a$ describes a strength of the repulsive potential near the origin. The last negative item in Eq.~\eqref{Potential1} guarantees that the sole minimum of the potential that is achieved at 
\begin{equation}\label{Rmin1}
r_{min}=2^{1/2}a^{1/4}r_0
\end{equation}
is zero \cite{Olendski3}. It is shown that for each quantum orbital the sum of the Shannon components and products of the Fisher (Onicescu) position and momentum informations (energies) do not depend on the uniform intensity since the change of one item is exactly compensated by the opposite variation with $B$ of the conjugate counterpart. In vein of our discussion above it means, in particular, that the total amount of information that one has about the position and momentum of the nanoparticle can not be altered by the uniform magnetic field. In the framework of this model, some analytic results can be derived and construed from physical point of view; among others, a simple expression for the position Fisher information, Eq.~\eqref{Fisher2_X}, manifests that it does not depend on the azimuthal quantum number and its product with the momentum component is a $B$-independent quantity, as is the multiplication of the two parts of the Onicescu energy too. It is deducted that the position Shannon entropy and the energy spectrum depend on the AB flux in a very similar way what can be used for deducing one of them from the knowledge about the second item. Possible generalization of the obtained results is also discussed.

\section{Formulation and Discussion}\label{sec2}
\subsection{Position and momentum waveforms}\label{sec2_1}
\begin{figure}
\centering
\includegraphics[width=0.7\columnwidth]{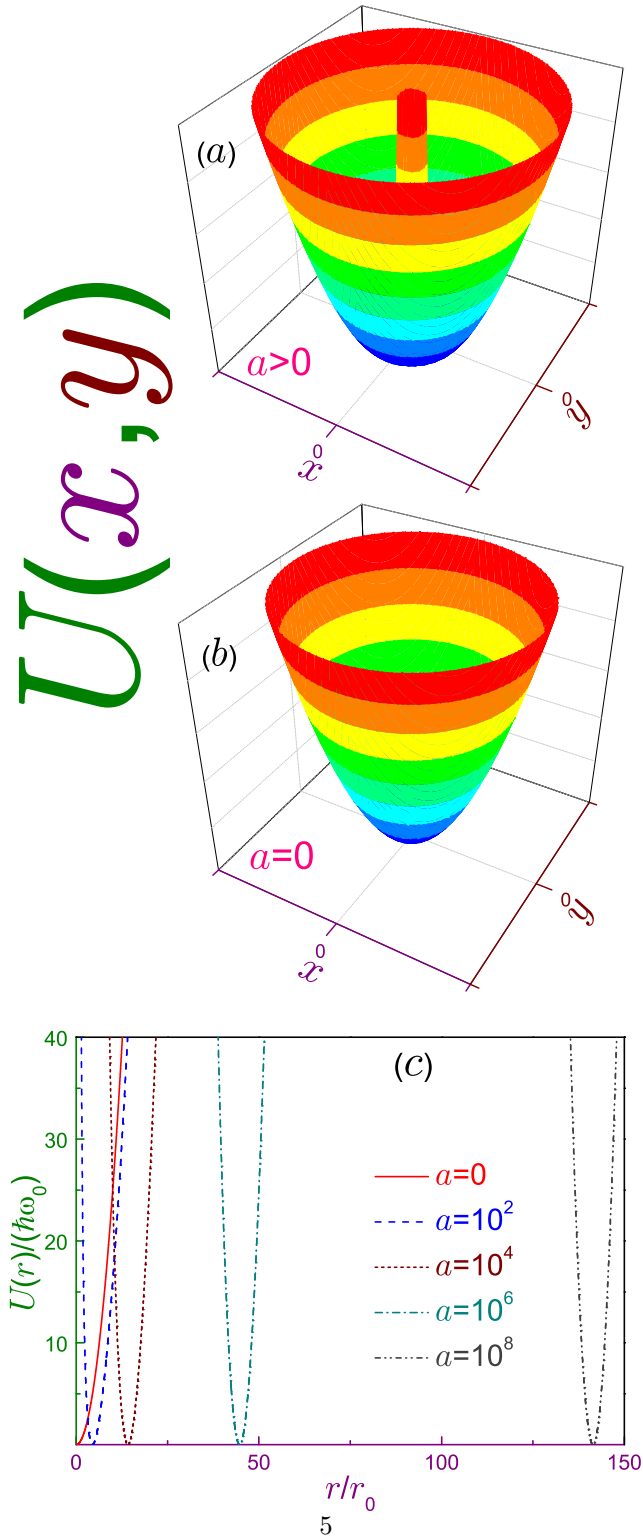}
\caption{\label{ProfileFig1}Potential profile $U(x,y)$ (in arbitrary units) as a function of the position in the $(x,y)$-plane for (a) the QR, $a>0$, and (b) QD, $a=0$. (c) Potential $U(r)$ (in units of $\hbar\omega_0$) as a function of the normalized radial distance $r/r_0$ for the QD (solid line), QR with $a=10^2$ (dashed curve), $a=10^4$ (dotted line), $a=10^6$ (dash-dotted dependence) and $a=10^8$ (dash-dot-dotted line).
} 
\end{figure}

We start our discussion by noting that the variation of the parameters of the potential from Eq.~\eqref{Potential1} allows one to describe different structures; first, as pointed out in the Introduction, for $a\omega_0\neq0$ it represents an isolated ring of the finite width and average radius from Eq.~\eqref{Rmin1}. Next, a model with $a=0$ corresponds to the quantum dot (QD) whereas an absence of the external boundary, $\omega_0=0$, portrays a quantum antidot \cite{Bogachek1} with its repulsive ability being proportional to the dimensionless parameter $a$. Moreover, the limit $\omega_0\rightarrow\infty$ with the requirement of the constant radius $r_{min}$ reduces the problem to the 1D ring whereas a suitable form for the 2D straight wire is given by the condition of the constant $\omega_0$ and unrestrictedly enlarging $r_{min}$. It is known that the QR with the small antidot strength describes a "thick" ring whereas the structure with the large $a$ is deemed as a "thin" annulus \cite{Olendski3}. A huge advantage of the theoretical consideration of Eq.~\eqref{Potential1} is due to the fact that it allows an exact (and quite simple and tractable, as is re-shown below) analytical solutions for the energy spectrum and position functions. Despite its simplicity, such representation of the potential correctly explains experimental data; in particular, it was successful in construing the beating effect in the oscillation pattern \cite{Liu1} and a magnitude of the persistent current in a single GaAs/Al$_x$Ga$_{1-x}$As loop \cite{Mailly1}. Fig.~\ref{ProfileFig1}(a) schematically depicts the azimuthally symmetric profile $U(r)$ in terms of the Cartesian components $x=r\cos\varphi_r$ and $y=r\sin\varphi_r$ for the QR and a comparison with its QD counterpart, window (b), explicitly shows that any non-zero antidot strength converts the geometry from the singly-connected configuration at $a=0$ to the doubly-connected one. Panel (c) demonstrates a transformation of the QD, $a=0$, through the "thick" ring (finite small and moderate $a$) to the "thin" one at $a\rightarrow\infty$.

In addition to the electrostatic potential, the 2D charged particle motion  is influenced by the magnetic field ${\bf B}_{tot}$ that is directed perpendicular to the $(x,y)$-plane, ${\bf B}_{tot}=B_{tot}{\bf k}$, and presents a superposition of the uniform component $B$ and extremely localized AB part \cite{Lewis1,Skarzhinskiy1,Bagrov1}:
\begin{equation}\label{Field1}
 B_{tot}=B+\phi_{AB}\delta(x)\delta(y)=B+\phi_{AB}\frac{\delta(r)}{\pi r},
\end{equation}
where $\delta(\xi)$ is a 1D $\delta$-function and $\phi_{AB}$ is the AB flux concentrated at the QR origin only.

For finding energies $E$ and position waveforms $\Psi({\bf r})$, one needs to address the 2D Schr\"{o}dinger equation
\begin{equation}\label{Schrodinger1}
\left[\frac{1}{2m^*}(-i\hbar{\bm\nabla}_{\bf r}+e{\bf A}_{tot})^2+U(r)\right]\Psi({\bf r})=E\Psi({\bf r})
\end{equation}
(here, $e$ is an absolute value of the electronic charge), that is most conveniently solved in the polar coordinates introduced above; in particular, in the symmetric gauge, the vector potential ${\bf A}_{tot}({\bf r})$ that describes the influence of the magnetic fields on a charged particle via the relation ${\bf B}_{tot}={\bm\nabla}_{\bf r}\times{\bf A}_{tot}$ has a nonzero azimuthal component only which depends solely on the distance from the center:
\begin{equation}\label{VectorPotential1}
{A_{tot}}_r=0,\quad {A_{tot}}_{\varphi_r}(r)=\frac{1}{2}Br+\frac{\phi_{AB}}{2\pi r}.
\end{equation}
Here, the first item in ${A_{tot}}_{\varphi_r}$ is due to the uniform intensity $B$ of the total field from Eq.~\eqref{Field1} and the AB flux yields the inverse-distance-like dependence in Eq.~\eqref{VectorPotential1}. Note that this zero-divergence potential satisfies the Stokes theorem:
\begin{equation}\label{Stokes1}
\oint_C{\bf A}_{tot}\cdot d{\bf l}=F_C,
\end{equation}
where $d{\bf l}$ is an elementary displacement along the closed arbitrary contour $C$ and $F_C$ is a net magnetic flux piercing it.

Then, as a result of representation from Eq.~\eqref{VectorPotential1}, the normalized, Eq.~\eqref{OrthoNormality1},
\begin{equation}\label{NormalizedPosition1}
\int_0^{2\pi}d\varphi_r\int_0^\infty dr r\Psi_{n'm'}^*\!\left(r,\varphi_r\right)\Psi_{nm}\!\left(r,\varphi_r\right)=\delta_{nn'}\delta_{mm'},
\end{equation}
position wave function is separated as a product of the angular and radial dependencies:
\begin{subequations}\label{PositionFunction1}
\begin{align}\label{PositionFinction1_1}
&\Psi_{nm}\!\left(r,\varphi_r\right)=\frac{1}{(2\pi)^{1/2}}e^{im\varphi_r}R_{nm}(r)
\intertext{with the latter one being expressed as:}
\label{PositionFinction1_2}
&R_{nm}(r)=\frac{1}{r_{eff}}\!\!\left[\frac{n!}{\Gamma(n\!+\!\lambda\!+\!1)}\right]^{1/2}\!\!\!\!\exp\!\left(\!-\frac{1}{4}\frac{r^2}{r_{eff}^2}\!\!\right)\!\left(\!\frac{1}{2}\frac{r^2}{r_{eff}^2}\!\!\right)^{\lambda/2}\!\!\!\!L_n^\lambda\!\left(\!\frac{1}{2}\frac{r^2}{r_{eff}^2}\!\!\right),
\end{align}
\end{subequations}
where $\Gamma(x)$ is $\Gamma$-function \cite{Abramowitz1}, $L_n^\alpha(x)$ is a generalized Laguerre polynomial  \cite{Abramowitz1},
\begin{subequations}\label{Quantities1}
\begin{align}\label{Quantities1_1}
r_{eff}&=\left(\frac{\hbar}{2m^*\omega_{eff}}\right)^{1/2}\\
\label{Quantities1_4}
\omega_{eff}&=\left(\omega_0^2+\frac{1}{4}\omega_c^2\right)^{1/2}
\intertext{and}
\label{Quantities1_5}
\omega_c&=\frac{eB}{m^*}
\intertext{is the cyclotron frequency,}
\label{Quantities1_2}
\lambda&=\left(m_\phi^2+a\right)^{1/2}\\
\label{Quantities1_3}
m_\phi&=m+\nu
\intertext{with $\nu$ being a dimensionless AB flux, i.e., the latter one is expressed in units of the elementary flux quantum $\phi_0=h/e$:}
\label{nu1}\nu&=\frac{\phi_{AB}}{\phi_0},
\end{align}
\end{subequations}
and $n=0,1,\ldots$, and $m=0,\pm1,\ldots$, are the principal and magnetic quantum numbers, respectively. Corresponding energy spectrum reads:
\begin{equation}\label{Energy0}
E_{nm}(a,\nu;\omega_c)=\hbar\omega_{e\!f\!f}\left(2n+\lambda+1\right)+\frac{1}{2}m_\phi\hbar\omega_c-\hbar\omega_0a^{1/2}.
\end{equation} 
For completeness, expressions for the persistent current $J_{nm}$ and magnetization $M_{nm}$ are given below too:
\begin{subequations}\label{CurrMagnet1}
\begin{align}\label{CurrMagnet1_Curr}
J_{nm}&\equiv-\frac{e}{h}\frac{\partial E}{\partial m}\equiv-\frac{\partial E}{\partial\phi_{AB}}=-\frac{e\omega_0}{2\pi}\left[\frac{m_\phi}{\lambda}\sqrt{1+\frac{1}{4}\left(\frac{\omega_c}{\omega_0}\right)^2}+\frac{1}{2}\frac{\omega_c}{\omega_0}\right]\\
\label{CurrMagnet1_Magn}
M_{nm}&\equiv-\frac{\partial E}{\partial B}=-\frac{e\hbar}{2m^*}\left[\frac{1}{2}\frac{\omega_c}{\omega_{eff}}(2n+\lambda+1)+m_\phi\right].
\end{align}
\end{subequations}
Note that Eqs.~\eqref{PositionFinction1_2}, \eqref{Energy0} and \eqref{CurrMagnet1} stay invariant under the transformation
\begin{equation}\label{Transformation1}
m\rightarrow m-1,\quad\nu\rightarrow\nu+1;
\end{equation}
accordingly, for analyzing the properties of these physical quantities, it is sufficient to study the AB flux inside the range
\begin{equation}\label{ABrange1}
-\frac{1}{2}\leq\nu\leq\frac{1}{2}.
\end{equation}
Observe also that at the zero uniform field, $B=0$, edges of this interval are characterized by the two infinite sets of the degeneracy of the levels:
\begin{subequations}\label{Degeneracy1}
\begin{align}\label{Degeneracy1_1}
E_{nm}\!\left(\!a,-\frac{1}{2};0\right)&=E_{n,-m+1}\!\left(\!a,-\frac{1}{2};0\right)\\
\label{Degeneracy1_2}
E_{nm}\!\left(\!a,\frac{1}{2};0\right)&=E_{n,-m-1}\!\left(\!a,\frac{1}{2};0\right).
\end{align}
\end{subequations}

For finding momentum quantum-information measures, one needs to evaluate first the corresponding waveform, which, according to Eq.~\eqref{Fourier1_1}, is:
\begin{equation}\label{MomentumFUnction1}
\Phi_{nm}({\bf k})\equiv\Phi_{nm}\!\left(k,\varphi_k\right)=\frac{1}{2\pi}\!\int_0^{2\pi}\!\!\!\!d\varphi_r\!\!\int_0^\infty\!\!\!\!drrR_{nm}(r)e^{i\left[m\varphi_r-kr\cos\left(\varphi_r-\varphi_k\right)\right]}.
\end{equation}
Azimuthal integration is carried out analytically what allows to represent the function $\Phi$ as a product of the angular and radial dependencies too:
\begin{subequations}\label{MomentumFunction2}
\begin{align}\label{MomentumFunction2_1}
&\Phi_{nm}\!\left(k,\varphi_k\right)=\frac{(-i)^m}{(2\pi)^{1/2}}e^{im\varphi_k}K_{nm}(k),
\intertext{where the latter factor is:}
\label{MomentumFunction2_2}
&K_{nm}(k)=r_{eff}\!\!\left[\frac{n!}{\Gamma(n\!+\!\lambda\!+\!1)}\right]^{1/2}\!\!\!\!\int_0^\infty\!\!\!e^{-z/2}z^{\lambda/2}L_n^\lambda(z)J_{|m|}\!\left(\!2^{1/2}r_{eff}kz^{1/2}\!\right)dz,
\end{align}
\end{subequations}
where $J_m(\xi)$ is $m$-th order Bessel function \cite{Abramowitz1}. Analytic expressions for the integrals of this type are missing in known literature \cite{Abramowitz1,Gradshteyn1,Prudnikov1,Brychkov1}; accordingly, below their direct numerical quadrature was employed. The only exception is the ground band, $n=0$, of the QD, $a=0$, without the AB flux, $\phi_{AB}=0$ (when, accordingly, $\lambda=|m|$):
\begin{subequations}\label{MomentumFunction3}
\begin{align}\label{MomentumFunction3_1}
\left.K_{0m}(k)\right|_{a=\nu=0}&=r_{eff}\frac{2^{|m|/2+1}}{(|m|!)^{1/2}}(r_{eff}k)^{|m|}\exp\!\left(-r_{eff}^2k^2\right);
\intertext{in particular:}
\label{MomentumFunction3_2}
\left.K_{00}(k)\right|_{a=\nu=0}&=2r_{eff}\exp\!\left(-r_{eff}^2k^2\right).
\end{align}
\end{subequations}
Comparison of the last formula with Eq.~\eqref{PositionFinction1_2} shows that these are position and momentum waveforms of the lowest-energy orbital of the 2D harmonic oscillator what leads, as will be shown below, to the saturation of the entropic uncertainty relation, Eq.~\eqref{EntropyInequality1}.

\begin{figure}
\centering
\includegraphics[width=\columnwidth]{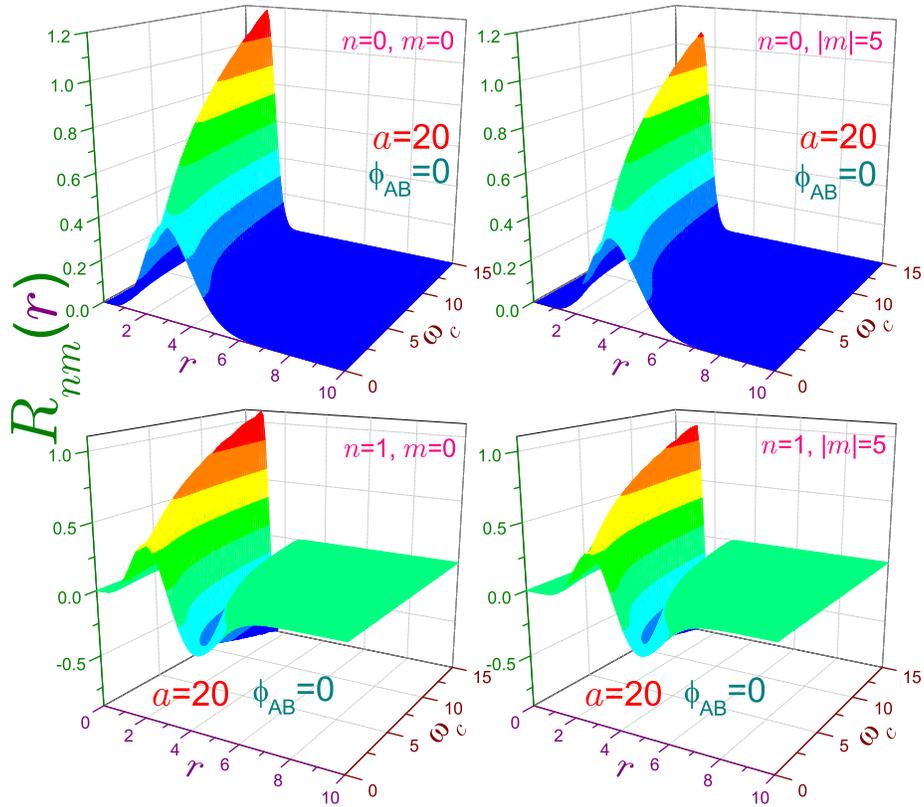}
\caption{\label{FunctionsR}Position radial waveforms $R_{nm}$ (in units of $r_0^{-1}$) as functions of the distance $r$ (measured in units of $r_0$) and cyclotron frequency $\omega_c$ (measured in units of $\omega_0$) at the antidot strength $a=20$ for several quantum numbers $n$ and $m$ denoted in each of the corresponding panels; namely, upper (lower) subplots are for the principal index $n=0$ ($n=1$) whereas left (right) ones depict the states with the azimuthal quantum number $m=0$ ($|m|=5$). AB flux is zero, $\phi_{AB}=0$. Note different ranges of the vertical axes in the upper and lower windows.
}
\end{figure}

\begin{figure}
\centering
\includegraphics[width=\columnwidth]{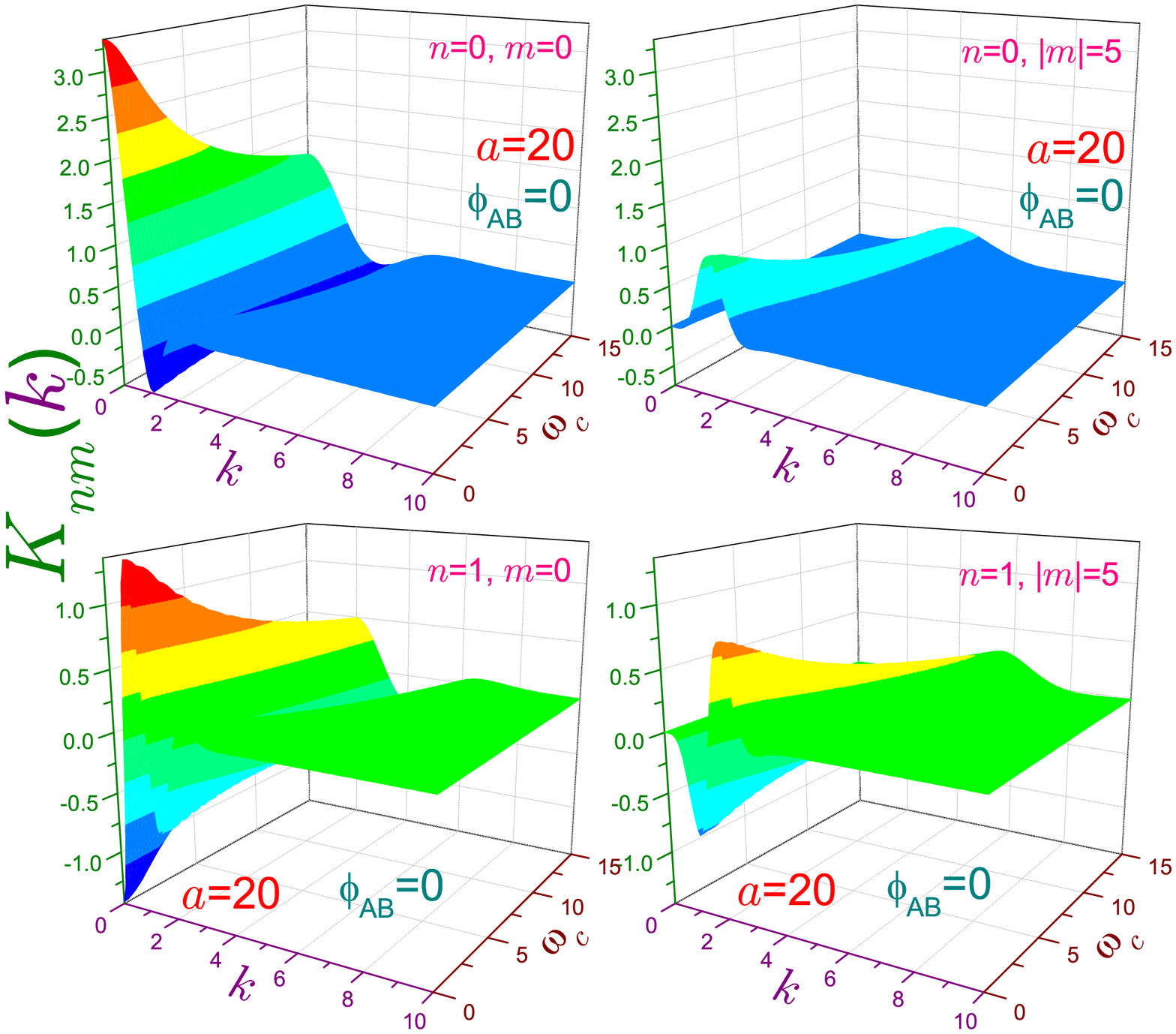}
\caption{\label{FunctionsK}Momentum radial waveforms $K_{nm}$ (in units of $r_0$) as functions of the wave vector $k$ (measured in units of $r_0^{-1}$) and cyclotron frequency $\omega_c$ (in units of $\omega_0$) for several quantum numbers $n$ and $m$ denoted in each of the corresponding panels. The same parameters as in Fig.~\ref{FunctionsR} are used. Upper and lower subplots have different ranges of their vertical axes.}
\end{figure}

Figs.~\ref{FunctionsR} and \ref{FunctionsK} depict position $R_{nm}(r)$ and momentum $K_{nm}(k)$ waveforms, respectively, as functions of the normalized cyclotron frequency $\omega_c/\omega_0$ with zero AB flux, $\phi_{AB}=0$, for several $n$ and $m$. As it follows from the general laws of quantum mechanics \cite{Landau1}, the principal index $n$ determines a number of nodes of the radial dependence $R_{nm}$ (excluding the points at the origin, $r=0$, and infinity, $r=\infty$) whereas the higher absolute values of the azimuthal index correspond to the location of the electric charge further away from the origin. Increasing field pushes the particle closer to the geometrical center of the ring with the simultaneous increase of the function $R_{nm}(r)$ extrema. In the wave vector domain, each dependence $K_{nm}(k)$ with $|m|\geq1$ is characterized by $n+1$ extrema whereas its $K_{n0}(k)$ counterpart has at $k=0$ an additional minimum or maximum with the largest amplitude. As it follows from Eq.~\eqref{MomentumFunction2_2}, in the limit of the infinitely strong magnetic intensities, the waveform $\Phi_{nm}({\bf k})$ turns to zero:
\begin{subequations}\label{FunctionLimits1}
\begin{align}\label{FunctionLimits1_1}
\Phi_{nm}({\bf k})&\xrightarrow[B\rightarrow\infty]{}0.
\intertext{This flattening of the radial parts of the momentum functions is clearly seen in Fig.~\ref{FunctionsK}. This also means that its Fourier transform $\Psi_{nm}({\bf r})$ tends in the same limit to the 2D $\delta$-function:}
\label{FunctionLimits1_2}
\Psi_{nm}({\bf r})&\xrightarrow[B\rightarrow\infty]{}\delta({\bf r}),
\end{align}
\end{subequations}
what physically means an extremely strong field-induced squeezing of the charge near the origin.

\begin{table*}[ht]
\caption{Field-independent quantum-information measures of the QR with $a=20$ for several values of $n$ and $|m|$}
\begin{tabular}{|c|c|c|c|c|c|c|}
\hline
\multicolumn{7}{|c|}{}\\
$n$&$\left|m\right|$&${S_\rho}_{nm}+{S_\gamma}_{nm}$&${I_\rho}_{nm}{I_\gamma}_{nm}$&${O_\rho}_{nm}{O_\gamma}_{nm}$&${CGL_\rho}_{nm}$&${CGL_\gamma}_{nm}$\\ \hline
\multirow{7}{*}{0}&0&5.0753&87.554&$1.4892\times10^{-2}$&1.1767&2.0253\\
&1&5.4924&55.782&$6.8465\times10^{-3}$&1.1764&1.4132\\
&2&5.8395&39.938&$4.2998\times10^{-3}$&1.1757&1.2567\\
&3&6.1478&31.492&$2.9983\times10^{-3}$&1.1748&1.1936\\
&4&6.4159&26.667&$2.2439\times10^{-3}$&1.1739&1.1689\\
&5&6.6452&23.724&$1.7701\times10^{-3}$&1.1731&1.1605\\
&10&7.3946&18.469&$8.3639\times10^{-4}$&1.1703&1.1629\\\hline
\multirow{7}{*}{1}&0&6.6206&$3.5866\times10^2$&$2.1896\times10^{-3}$&1.1809&1.3914\\
&1&6.6648&$3.1596\times10^2$&$2.0706\times10^{-3}$&1.1802&1.3761\\
&2&6.8000&$2.7022\times10^2$&$1.6819\times10^{-3}$&1.1782&1.2817\\
&3&6.9821&$2.3612\times10^2$&$1.3176\times10^{-3}$&1.1755&1.2075\\
&4&7.1741&$2.1257\times10^2$&$1.0493\times10^{-3}$&1.1727&1.1678\\
&5&7.3559&$1.9640\times10^2$&$8.5991\times10^{-4}$&1.1700&1.1505\\
&10&8.0174&$1.6296\times10^2$&$4.3886\times10^{-4}$&1.1611&1.1465\\\hline
\multirow{7}{*}{2}&0&7.1226&$7.5777\times10^2$&$1.9495\times10^{-3}$&1.1987&2.0162\\
&1&7.2577&$6.6632\times10^2$&$1.2081\times10^{-3}$&1.1976&1.4315\\
&2&7.3559&$6.0814\times10^2$&$1.0250\times10^{-3}$&1.1946&1.3432\\
&3&7.4906&$5.6355\times10^2$&$8.4033\times10^{-4}$&1.1907&1.2641\\
&4&7.6455&$5.2952\times10^2$&$6.8589\times10^{-4}$&1.1865&1.2089\\
&5&7.8007&$5.0393\times10^2$&$5.7048\times10^{-4}$&1.1825&1.1783\\
&10&8.4031&$4.4294\times10^2$&$3.0205\times10^{-4}$&1.1687&1.1530\\\hline
\multirow{7}{*}{3}&0&7.6604&$1.2849\times10^3$&$8.5659\times10^{-4}$&1.2174&1.4935\\
&1&7.7449&$1.1852\times10^3$&$7.0509\times10^{-4}$&1.2161&1.3393\\
&2&7.7990&$1.0950\times10^3$&$6.6512\times10^{-4}$&1.2124&1.3376\\
&3&7.8885&$1.0283\times10^3$&$5.8556\times10^{-4}$&1.2075&1.2930\\
&4&8.0102&$9.7956\times10^2$&$4.9556\times10^{-4}$&1.2023&1.2414\\
&5&8.1424&$9.4329\times10^2$&$4.1998\times10^{-4}$&1.1972&1.2058\\
&10&8.6942&$8.5450\times10^2$&$2.3004\times10^{-4}$&1.1794&1.1641\\\hline
\multirow{7}{*}{4}&0&7.9610&$1.9400\times10^3$&$8.5694\times10^{-4}$&1.2353&1.9889\\
&1&8.0892&$1.7951\times10^3$&$5.3022\times10^{-4}$&1.2337&1.4007\\
&2&8.1491&$1.6946\times10^3$&$4.8088\times10^{-4}$&1.2296&1.3533\\
&3&8.2143&$1.6164\times10^3$&$4.3744\times10^{-4}$&1.2239&1.3201\\
&4&8.3113&$1.5566\times10^3$&$3.8042\times10^{-4}$&1.2178&1.2713\\
&5&8.4247&$1.5111\times10^3$&$3.2768\times10^{-4}$&1.2119&1.2324\\
&10&8.9326&$1.3959\times10^3$&$1.8494\times10^{-4}$&1.1906&1.1766\\\hline
\multirow{7}{*}{5}&0&8.3096&$2.7231\times10^3$&$4.8473\times10^{-4}$&1.2520&1.5729\\
&1&8.3891&$2.5665\times10^3$&$3.8797\times10^{-4}$&1.2502&1.3650\\
&2&8.4454&$2.4338\times10^3$&$3.6107\times10^{-4}$&1.2456&1.3489\\
&3&8.4932&$2.3357\times10^3$&$3.3947\times10^{-4}$&1.2393&1.3371\\
&4&8.5703&$2.2626\times10^3$&$3.0292\times10^{-4}$&1.2325&1.2959\\
&5&8.6676&$2.2074\times10^3$&$2.6506\times10^{-4}$&1.2259&1.2566\\
&10&9.1367&$2.0664\times10^3$&$1.5385\times10^{-4}$&1.2018&1.1893\\\hline
\end{tabular}
\label{Table1}
\end{table*}

\subsection{Measures in uniform magnetic field $\bf B$}\label{sec2_2}
Knowledge of the waveforms $\Psi_{nm}({\bf r})$ and $\Phi_{nm}({\bf k})$ and, accordingly, of the corresponding densities $\rho_{nm}({\bf r})$ and $\gamma_{nm}({\bf k})$ paves the way to calculating quantum-information measures; for example, position and momentum Shannon entropies are:
\begin{subequations}\label{Shannon2}
\begin{align}
&S_{\rho_{nm}}=2\ln r_{eff}+\ln2\pi-\ln\frac{n!}{\Gamma(n+\lambda+1)}\nonumber\\
\label{Shannon2_X}
&-\frac{n!}{\Gamma(n+\lambda+1)}\int_0^\infty\!e^{-z}z^\lambda L_n^\lambda(z)^2\ln\left(e^{-z}z^\lambda L_n^\lambda(z)^2\right)\!dz\\
&S_{\gamma_{nm}}=-2\ln r_{eff}+\ln2\pi-\ln\frac{n!}{\Gamma(n+\lambda+1)}\nonumber\\
\label{Shannon2_K}
&-\frac{n!}{\Gamma(n+\lambda+1)}\int_0^\infty\!d\xi\xi\left[\int_0^\infty\!e^{-z/2}z^{\lambda/2}L_n^\lambda(z)J_{|m|}\!\left(2^{1/2}\xi z^{1/2}\right)\!dz\right]^2\nonumber\\
&\times\ln\!\left(\left[\int_0^\infty\!e^{-z/2}z^{\lambda/2}L_n^\lambda(z)J_{|m|}\!\left(2^{1/2}\xi z^{1/2}\right)\!dz\right]^2\right).
\intertext{These expressions manifest that the whole dependence of the position (momentum) entropy on the magnetic intensity $B$ is reflected in the negative (positive) term $\ln\!\left(1+\frac{1}{4}\frac{\omega_c^2}{\omega_0^2}\right)$ only what means that their sum is a field-independent quantity:}
&S_{\rho_{nm}}+S_{\gamma_{nm}}=2\ln2\pi-2\ln\frac{n!}{\Gamma(n+\lambda+1)}\nonumber\\
&-\frac{n!}{\Gamma(n+\lambda+1)}\int_0^\infty\!e^{-z}z^\lambda L_n^\lambda(z)^2\ln\left(e^{-z}z^\lambda L_n^\lambda(z)^2\right)\!dz\nonumber\\
&-\frac{n!}{\Gamma(n+\lambda+1)}\int_0^\infty d\xi\xi\left[\int_0^\infty\!e^{-z/2}z^{\lambda/2}L_n^\lambda(z)J_{|m|}\!\left(2^{1/2}\xi z^{1/2}\right)\!dz\right]^2\nonumber\\
\label{Shannon2_T}
&\times\ln\!\left(\left[\int_0^\infty\!e^{-z/2}z^{\lambda/2}L_n^\lambda(z)J_{|m|}\!\left(2^{1/2}\xi z^{1/2}\right)\!dz\right]^2\right).
\end{align}
\end{subequations}
Thus, an amount of the total information that is available simultaneously about position and motion of the nanoparticle can not be altered by the uniform magnetic fields. For the strong $B$, position (momentum) Shannon entropy decreases (increases) as $\mp\ln\omega_c$, what is consistent with the interpretation of the entropies as measures of uncertainty: in this regime, we know more (less) about particle location (momentum), as it follows from Figs.~\ref{FunctionsR} and \ref{FunctionsK}; accordingly, the corresponding measure gets smaller (larger) tending to the negative (positive) infinity. However, the loss at the growing fields of the information about the electron momentum is exactly compensated by the extra knowledge we acquire about its position. As a result, the sum of the two entropies does not depend on $B$. For the ground band, $n=0$, the position component can be calculated analytically:
\begin{subequations}\label{Shannon3}
\begin{align}\label{Shannon3_1}
{S_\rho}_{0m}&=2\ln r_{eff}+1+\ln2\pi+\ln\Gamma(\lambda+1)+\lambda[1-\psi(\lambda+1)],
\intertext{where $\psi(x)=d[\ln\Gamma(x)]/dx=\Gamma'(x)/\Gamma(x)$ is psi, or digamma function, which is a derivative of the logarithm of the $\Gamma$-function \cite{Abramowitz1}. Asymptotic form of this expression}
\label{Shannon3_2}
{S_\rho}_{0m}&\rightarrow2\ln r_{eff}+\frac{3}{2}\ln2\pi+\frac{1}{2}+\frac{1}{2}\ln\lambda,\quad\lambda\rightarrow\infty,
\intertext{shows that at the very large magnitudes of the magnetic indices the entropy ${S_\rho}_{0m}$ increases as $\frac{1}{2}\ln|m|$. In the opposite limit of the azimuthally symmetric state, $m=0$, the QD ($a=0$) entropy without the AB flux, $\phi_{AB}=0$, is:}
\label{Shannon3_3}
\left.{S_\rho}_{00}\right|_{a=\nu=0}&=2\ln r_{eff}+1+\ln\pi+\ln2.
\end{align}
\end{subequations}
On the other hand, with the help of the function from Eq.~\eqref{MomentumFunction3_1} one evaluates the momentum entropy as
\begin{subequations}\label{Shannon4}
\begin{align}\label{Shannon4_1}
\left.{S_\gamma}_{0m}\right|_{a=\nu=0}&=-2\ln r_{eff}+1+\ln\!\left(\frac{|m|!}{2^{|m|+1}}\pi\!\right)-2^{|m|+1}|m|\psi(|m|+1);
\intertext{in particular:}
\label{Shannon4_2}
\left.{S_\gamma}_{00}\right|_{a=\nu=0}&=-2\ln r_{eff}+1+\ln\pi-\ln2.
\end{align}
\end{subequations}
Comparing Eqs.~\eqref{Shannon3_3} and \eqref{Shannon4_2}, one sees that the QD ground state does saturate the entropic uncertainty relation, Eq.~\eqref{EntropyInequality1}, regardless of the magnetic intensity $B$. As mentioned above, this takes place since the corrresponding position and momentum dependencies are those of the 2D harmonic oscillator. 

\begin{figure}
\centering
\includegraphics[width=\columnwidth]{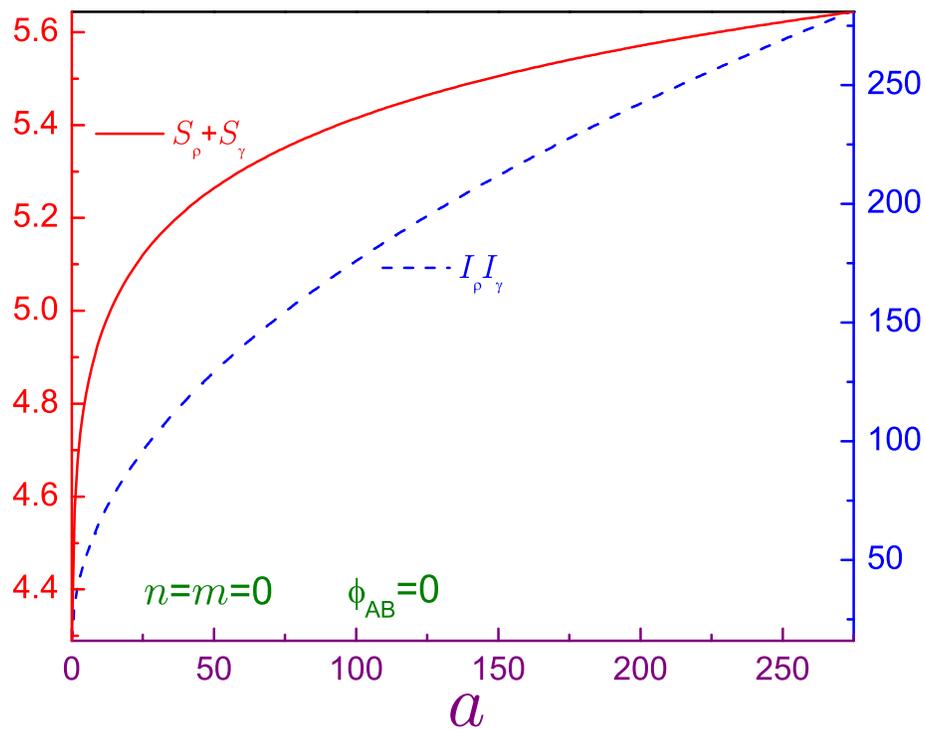}
\caption{\label{ShannonFisherFig1}Sum of the entropies ${S_\rho}_{00}+{S_\gamma}_{00}$ (solid line, left axis) and product of the Fisher informations ${I_\rho}_{00}{I_\gamma}_{00}$ (dashed line, right axis) as functions of the QR "thickness" $a$. The dependencies are universal in a sense that they do not depend on the uniform magnetic field $B$.}
\end{figure}

\begin{figure}
\centering
\includegraphics[width=\columnwidth]{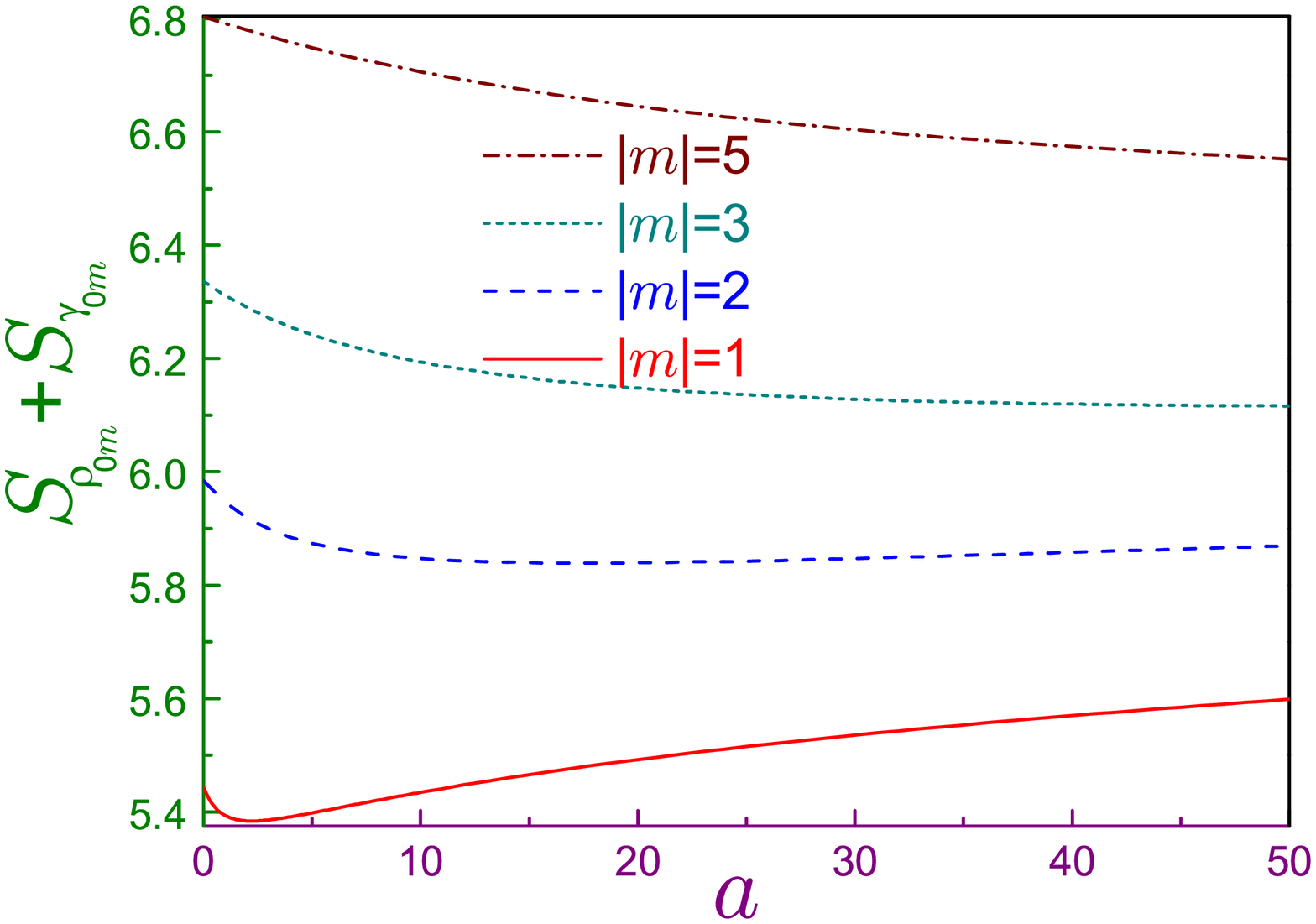}
\caption{\label{ShannonFig2}Sum ${S_\rho}_{0m}+{S_\gamma}_{0m}$ for several quantum numbers $m$ as a function of the antidot strength $a$ where solid line is for $|m|=1$, dashed curve -- for $|m|=2$, dotted dependence -- for $|m|=3$, and dash-dotted line is for $|m|=5$.}
\end{figure}

Solid line in Fig.~\ref{ShannonFisherFig1} depicts a variation of the field-indepdendent quantity ${S_\rho}_{00}+{S_\gamma}_{00}$ with the strength of the antidot $a$ for the zero AB flux, $\phi_{AB}=0$. At the small $a$, corresponding to the "thick" ring \cite{Olendski3}, the sum increases quite rapidly from its threshold QD value $2(1+\ln\pi)=4.2894\ldots$ whereas as the annulus  gets thinner \cite{Olendski3}, the growth of the total entropy slows down. Different behaviour is  observed for the sum of the two entropies of the azimuthally asymmetric states, $m\neq0$; namely, as Fig.~\ref{ShannonFig2} demonstrates, it has a pronounced minimum whose location on the $a$ axis for the greater quantum indices $|m|$ is shifted to the right with the simultaneous broadening of this extremum. Entropy dependencies depicted in Figs.~\ref{ShannonFisherFig1} and \ref{ShannonFig2} for the ground band, $n=0$, are qualitatively repeated for the orbitals with the nonzero principal numbers $n$.

Table~\ref{Table1} lists the sums ${S_\rho}_{nm}+{S_\gamma}_{nm}$ for $n$ ranging from zero to five and $|m|=0..5$ and $10$ for the QR with $a=20$. It is seen that the total entropy is an increasing function of both the principal as well as magnetic quantum numbers.

A remarkable property of the position Fisher information
\begin{equation}\label{FisherPos1}
{I_\rho}_{nm}=\frac{8}{r_{eff}^2}\frac{n!}{\Gamma(n+\lambda+1)}\!\!\int_0^\infty\!\!\!\!e^{-z}z^{\lambda+1}\!\left[\frac{1}{2}\!\left(\frac{\lambda}{z}-1\right)L_n^\lambda(z)-L_{n-1}^{\lambda+1}(z)\right]^2\!\!dz
\end{equation}
is the fact that it does not depend on the azimuthal quantum number and the AB flux:
\begin{subequations}\label{Fisher2}
\begin{align}\label{Fisher2_X}
&{I_\rho}_{nm}=\frac{4n+2}{r_{eff}^2},
\intertext{as a direct evaluation \cite{Gradshteyn1,Prudnikov1} of Eq.~\eqref{FisherPos1} reveals. Momentum Fisher information reads:}
\label{Fisher2_K}
&{I_\gamma}_{nm}=\!r_{eff}^2\frac{8n!}{\Gamma(n\!+\!\lambda\!+\!1)}\!\int_0^\infty\!\!\!d\xi\xi\!\left[\int_0^\infty\!\!\!\!e^{-z/2}z^{(\lambda+1)/2}L_n^\lambda(z)J_{|m|}'\!\left(\!2^{1/2}\xi z^{1/2}\right)\!dz\right]^2\!\!\!\!,
\intertext{what makes the product of the two, similar to the sum of the Shannon entropies, a field-independent quantity:}
\label{Fisher2_T}
&{I_\rho}_{nm}{I_\gamma}_{nm}=\!\frac{8(4n+2)n!}{\Gamma(n\!+\!\lambda\!+\!1)}\!\int_0^\infty\!\!\!d\xi\xi\!\left[\int_0^\infty\!\!\!\!e^{-z/2}z^{(\lambda+1)/2}L_n^\lambda(z)J_{|m|}'\!\left(\!2^{1/2}\xi z^{1/2}\right)\!dz\right]^2\!\!\!\!.
\end{align}
\end{subequations}
One can interpret the $m$-independence of the position Fisher information by the fact that this quantum number determines a spatial localization of the particle but not a number of oscillations that are described by the principal index $n$, as stated above. Increasing magnetic intensity sharpens waveforms $R_{nm}(r)$, as Fig.~\ref{FunctionsR} vividly demonstrates, what leads to the linear growth of $I_\rho$ with $\omega_{eff}$ that at the high $B$ reduces to the linear dependence of the position component on the field. Simultaneously, the slope of the variation of the momentum functions $K_{nm}(k)$ becomes more gentle approaching in the limit $B\rightarrow\infty$ the (zero) uniform $k$-independent value, Eq.~\eqref{FunctionLimits1_1}, what means a vanishing of the momentum Fisher information. However, the product of the two stays the same since the increasing intensity  of the oscillations of the position waveform $\Psi_{nm}({\bf r})$ is exactly compensated by the flattening of its momentum counterpart. Using Eq.~\eqref{MomentumFunction3_1}, one derives the momentum Fisher information of the ground band of the QD as
\begin{equation}\label{Fisher3}
\left.{I_\gamma}_{0m}\right|_{a=\nu=0}=8r_{eff}^2,
\end{equation}
which is again a $m$-independent measure. Dashed line in Fig.~\ref{ShannonFisherFig1} shows that the quantity ${I_\rho}_{00}{I_\gamma}_{00}$ increases with the transformation of the QR from the 'thick' to the 'thin' geometry. Our results also reveal that for the greater $|m|$ a slope of the ${I_\rho}_{0m}{I_\gamma}_{0m}$ dependence on $a$ does decrease. In addition, an analysis of the products ${I_\rho}_{nm}{I_\gamma}_{nm}$ that are provided in Table~\ref{Table1} uncovers that it increases with the quantum number $n$ and gets smaller for the growing $|m|$.

\begin{figure}
\centering
\includegraphics[width=\columnwidth]{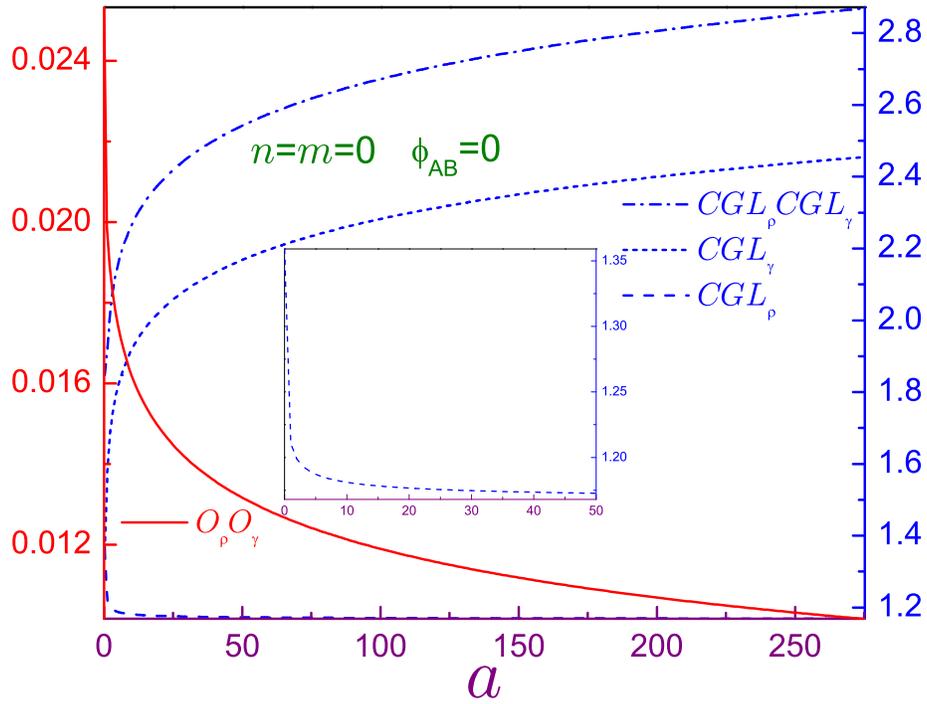}
\caption{\label{OnicescuFig1}Product ${O_\rho}_{00}{O_\gamma}_{00}$ (solid line, left axis) and complexities ${CGL_\rho}_{00}$ (dashed line, right axis), ${CGL_\gamma}_{00}$ (dotted line, right axis) and their product ${CGL_\rho}_{00}{CGL_\gamma}_{00}$ (dash-dotted line, right axis) as functions of the QR "thickness" $a$. Inset enlarges a dependence of the position complexity at the small and moderate $a$.}
\end{figure}

Expressions for the position
\begin{subequations}\label{Onicescu2}
\begin{align}\label{Onicescu2_X}
{O_\rho}_{nm}&=\frac{1}{r_{eff}^2}\frac{1}{2\pi}\left[\frac{n!}{\Gamma(n+\lambda+1)}\right]^2\int_0^\infty e^{-2z}z^{2\lambda}\left[L_n^\lambda(z)\right]^4dz
\intertext{and momentum}
\label{Onicescu2_K}
{O_\gamma}_{nm}&=\!\!\frac{r_{eff}^2}{2\pi}\!\!\left[\frac{n!}{\Gamma(n\!+\!\lambda\!+\!1)}\right]^2\!\!\!\int_0^\infty\!\!\!d\xi\xi\!\left[\int_0^\infty\!\!\!\!e^{-z/2}z^{\lambda/2}L_n^\lambda(z)J_{|m|}\!\left(\!2^{1/2}\xi z^{1/2}\right)\!dz\right]^4
\end{align}
\end{subequations}
QR Onicescu energies demonstrate that their dependencies on the uniform field are the same as for the Fisher informations what makes the product ${O_\rho}_{nm}{O_\gamma}_{nm}$ a $B$-independent quantity again. A growth (lessening) with the field of the position (momentum) component is explained by the increasing (decreasing) nonuniformity of the function $R_{nm}(r)$ [$K_{nm}(k)$] in this regime. Similar to the previous functionals, the variation of one measure is exactly counterbalanced by the change in the opposite direction of the second fellow in such a way that their product stays intact by the uniform field. Data in Table~\ref{Table1} show that ${O_\rho}_{nm}{O_\gamma}_{nm}$ is a decreasing function of either of the quantum indices. For $n=0$, Eq.~\eqref{Onicescu2_X} yields:
\begin{subequations}\label{Onicescu3}
\begin{align}\label{Onicescu3_1}
{O_\rho}_{0m}&=\frac{1}{r_{eff}^2}\frac{1}{4\pi^{3/2}}\frac{\Gamma(\lambda+\frac{1}{2})}{\Gamma(\lambda+1)},
\intertext{and its asymptote}
\label{Onicescu3_2}
{O_\rho}_{0m}&\rightarrow\frac{1}{r_{eff}^2}\frac{1}{4\pi^{3/2}}\frac{1}{\lambda^{1/2}},\quad\lambda\rightarrow\infty,
\end{align}
\end{subequations}
shows that at the large $|m|$ the position Onicescu energy fades as $|m|^{-1/2}$. For the QD without the AB flux, its momentum counterpart reads as
\begin{subequations}\label{Onicescu4}
\begin{align}\label{Onicescu4_1}
\left.{O_\gamma}_{0m}\right|_{a=\nu=0}&=\frac{r_{eff}^2}{\pi}\frac{(2|m|)!}{2^{2|m|}(|m|!)^2}
\intertext{with its limit at the huge $|m|$ being}
\label{Onicescu4_2}
\left.{O_\gamma}_{0m}\right|_{a=\nu=0}&\rightarrow\frac{r_{eff}^2}{\pi^{3/2}}\frac{1}{|m|^{1/2}},\quad|m|\rightarrow\infty.
\end{align}
\end{subequations}
Solid line in Fig.~\ref{OnicescuFig1} exemplifies that the field-independent product ${O_\rho}_{00}{O_\gamma}_{00}$ smoothly decreases as the QR becomes thinner; however, its $m\neq0$ companions exhibit the maximum in their $a$ dependence, which at the larger $|m|$ is shifted further to the right with its simultaneous broadening (not shown here). These features are qualitatively reproduced for the $n\neq0$ orbitals.

Comparing Eqs.~\eqref{Shannon2} and \eqref{Onicescu2}, one sees that both position and momentum parts of the complexity $CGL$ do not depend on the uniform field either. Their values are provided in the corresponding columns of Table~\ref{Table1}. As expected, each of them is greater than unity. Dependence of the $n=0$, $m=0$ complexities on the QAD strength is shown in Fig.~\ref{OnicescuFig1} by the dashed, dotted and dash-dotted lines for $CGL_\rho$, $CGL_\gamma$ and their product, respectively, under the assumption that the AB flux is zero. For the QD, both position and momentum complexities have the value of $e/2=1.3591\ldots$ with their product $e^2/4=1.8472\ldots$. The former one quite precipitously decreases as $a$ gets stronger approaching at $a\rightarrow\infty$ the limit of $(e/2)^{1/2}=1.1658\ldots$; however, the growth of the momentum part guarantees that their product smoothly increases as the QR becomes thinner.

\subsection{AB measures}\label{sec2_3}
\begin{figure}
\centering
\includegraphics[width=\columnwidth]{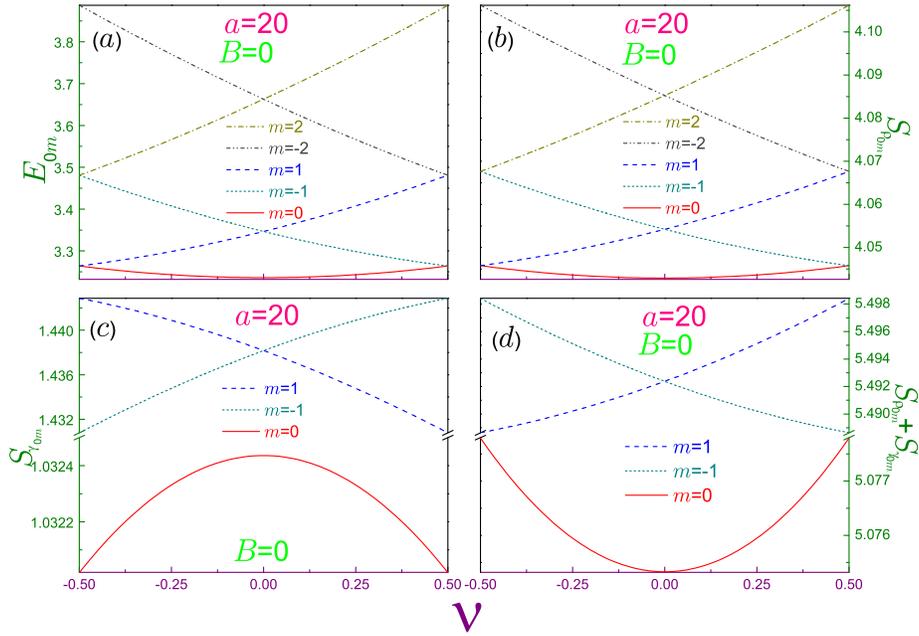}
\caption{\label{ShannonFig3}(a) Energy spectrum $E_{0m}$ (in units of $\hbar\omega_0$), (b) position ${S_\rho}_{0m}$, (c) momentum ${S_\gamma}_{0m}$ Shannon entropies and (d) their sum ${S_\rho}_{0m}+{S_\gamma}_{0m}$ as functions of the normalized AB flux $\nu$ for several magnetic quantum numbers $m$ denoted in each of the corresponding subplot. Antidot strength is $a=20$, magnetic filed is zero, $B=0$, and radius $r_0$ is assumed to be equal to unity. Note vertical axes breaks in panels (c) from 1.0325 to 1.4308 and (d) from 5.0778 to 5.4886 and different scales above and below the break.}
\end{figure}

Turning to the discussion of the measures' dependence on the AB flux, first let us write Taylor expansions of the position Shannon entropy, Eq.~\eqref{Shannon3_1}, and Onicescu energy, Eq.~\eqref{Onicescu3_1}, with respect to the AB parameter $\nu$ for the azimuthally symmetric orbital, $m=0$:
\begin{align}
{S_\rho}_{00}(\nu)&=2\ln r_{eff}+\ln2\pi+1+\ln\Gamma\!\left(a^{1/2}+1\right)+a^{1/2}\!\!\left[1-\psi\!\left(a^{1/2}\right)\right]\nonumber\\
\label{Shannon5}
&+\frac{1}{2}\left[\frac{1}{a^{1/2}}+\frac{1}{a}-\psi^{(1)}(a)\right]\nu^2\\
\label{Onicescu5}
{O_\rho}_{00}(\nu)&=\frac{1}{r_{eff}^2}\frac{1}{4\pi^{3/2}}\frac{\Gamma\!\left(a^{1/2}\!+\!\frac{1}{2}\right)}{\Gamma\!\left(a^{1/2}\!+\!1\right)}\!\left(\!1\!-\!\frac{1}{2a^{1/2}}\!\!\left[\psi\!\left(\!a^{1/2}\!+\!1\!\right)\!-\!\psi\!\!\left(\!a^{1/2}\!+\!\frac{1}{2}\right)\right]\!\!\nu^2\!\right),
\end{align}
where the series have been truncated at the first nonvanishing power of the flux and $\psi^{(n)}(x)=d^n\psi(x)/dx^n$, $n=1,2,\ldots$ is a polygamma function \cite{Abramowitz1}. It can be shown that the expression in the second square brackets of the right-hand side of Eq.~\eqref{Shannon5} is always positive what makes the entropy of the corresponding level a convex function of the AB flux. On the other hand, since $\psi(x)$ is an increasing function of its positive argument \cite{Abramowitz1}, Onicescu energy ${O_\rho}_{00}$ is a concave function of $\nu$. At the quite large $a$, utilizing properties of the polygamma function \cite{Abramowitz1}, one has:
\begin{align}\label{Shannon6}
{S_\rho}_{00}(\nu)&=2\ln r_{eff}+\frac{3}{2}\ln2\pi+\frac{1}{2}+\frac{1}{4}\ln a+\frac{1}{4a}\!\left(1-\frac{1}{3a^{1/2}}\right)\nu^2,\,a\gg1,\\
\label{Onicescu6}
{O_\rho}_{00}(\nu)&=\frac{1}{r_{eff}^2}\frac{1}{4\pi^{3/2}a^{1/4}}\!\left[1\!-\!\frac{1}{4a}\!\left(1-\frac{1}{4a^{1/2}}\right)\nu^2\right],\quad a\gg1.
\end{align}
These equations manifest that the opposite directions of the ${S_\rho}_{00}-\nu$ and ${O_\rho}_{00}-\nu$ characteristics have almost the same magnitude of their inclines. Comparing them with the Taylor expansion of the energy, Eq.~\eqref{Energy0}, at the zero uniform magnetic field:
\begin{equation}\label{Energy1}
E_{n0}(a,\nu;0)=\hbar\omega_0\!\left(\!2n+1+\frac{1}{2a^{1/2}}\nu^2\right),
\end{equation}
one sees that both the entropy and (normalized in units of $\hbar\omega_0$) energy are the AB convex functions with, however, different slopes.

Panels (a) and (b) of Fig.~\ref{ShannonFig3} provide a comparative analysis of the energy spectrum and position Shannon entropy dependencies on the AB field at the zero uniform component, $B=0$. A very strong similarity between the two is clearly seen; for example, for the ground level, $m=0$, if both $E_{00}$ and ${S_\rho}_{00}$ are appropriately scaled, the two lines practically coincide with each other. Thus, for evaluating the persistent current, Eq.~\eqref{CurrMagnet1_Curr}, one can use, in addition to (or instead of) the speed of change of the energy levels, the corresponding rate of variation of the associated position Shannon entropies multiplied by the normalizing coefficient $-e\omega_0a^{1/2}/\pi$. Of course, for getting the information on the persistent current, one can also use the Onicescu energy too but the related factor will have the opposite sign (and different value) as that for the entropy. For the inverse problem, measuring the current $J_{nm}(\nu)$, one can deduce what are the Shannon entropy and Onicescu energy of the corresponding orbital. Observe that ${S_\rho}_{nm}$, similar to the energy spectrum, persistent current and magnetization, stays invariant under the transformation from Eq.~\eqref{Transformation1}. The same is true for the position Onicescu energy. In addition, the degeneracy of the energies expressed by Eqs.~\eqref{Degeneracy1}, is inherited by ${S_\rho}_{nm}$ too. However, as Eqs.~\eqref{Shannon2_K}, \eqref{Fisher2_K} and \eqref{Onicescu2_K} show, it is not the case for the momentum components of all three measures. This remarkable difference is traced back to the corresponding waveforms, Eqs.~\eqref{PositionFinction1_2} and \eqref{MomentumFunction2_2}, respectively. Momentum Shannon entropies are depicted in Fig.~\ref{ShannonFig3}(c). It is seen that ${S_\gamma}_{00}$ is a concave function of the AB flux and that at $\nu=-1/2$ ($\nu=1/2$) it does not turn into ${S_\gamma}_{0,+1}$ $\left({S_\gamma}_{0,-1}\right)$, as just described above. For all other orbitals, the direction of change of the momentum Shannon entropy is opposite to that of its position counterpart too but since the relative magnitude of the former variation is smaller, the sum of the two entropies repeats as a function of $\nu$ the behaviour of the position component, as panel (d) of Fig.~\eqref{ShannonFig3} demonstrates. Obviously, the invariance of ${S_\rho}_{nm}+{S_\gamma}_{nm}$ under the transformation from Eq.~\eqref{Transformation1} is not conserved either.

\begin{figure}
\centering
\includegraphics[width=\columnwidth]{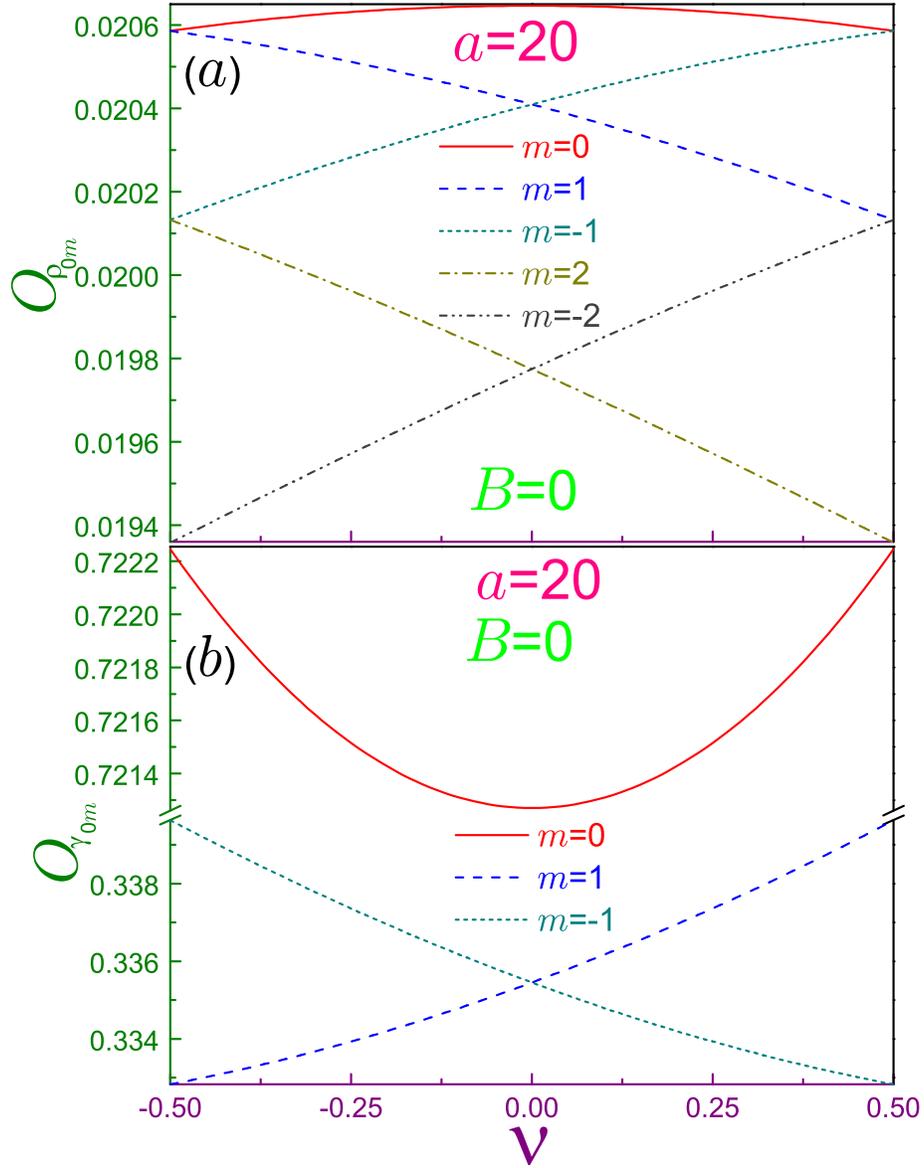}
\caption{\label{OnicescuFig2} Onicescu (a) position ${O_\rho}_{0m}$ and  (b) momentum ${O_\gamma}_{0m}$ energies as functions of the normalized AB flux $\nu$ for several magnetic quantum numbers $m$ denoted in each of the corresponding subplot. The same parameters as in Fig.~\ref{ShannonFig3} are used. Note vertical axis break in panel (b) from 0.33964 to 0.72126 and different scales above and below the break.}
\end{figure}

Panel (a) of Fig.~\ref{OnicescuFig2} that exhibits position Onicescu energy confirms our analytic result about the concavity of the function ${O_\rho}_{00}(\nu)$, Eq.~\eqref{Onicescu6}. Similar to this orbital, for any other level the direction of change of ${O_\rho}_{nm}$ is opposite to that of the position entropy too, cf. Figs.~\ref{ShannonFig3}(b) and \ref{OnicescuFig2}(a). As was the Shannon case, the deviation with $\nu$ of the  momentum Onicescu component is inverse to that of the position one; in particular, as window (b) of Fig.~\ref{OnicescuFig2} shows, ${O_\gamma}_{00}(\nu)$ is a convex function of the flux; however, the product ${O_\rho}_{00}(\nu){O_\gamma}_{00}(\nu)$ demonstrates a concave behaviour again (not shown here).

\begin{figure}
\centering
\includegraphics[width=\columnwidth]{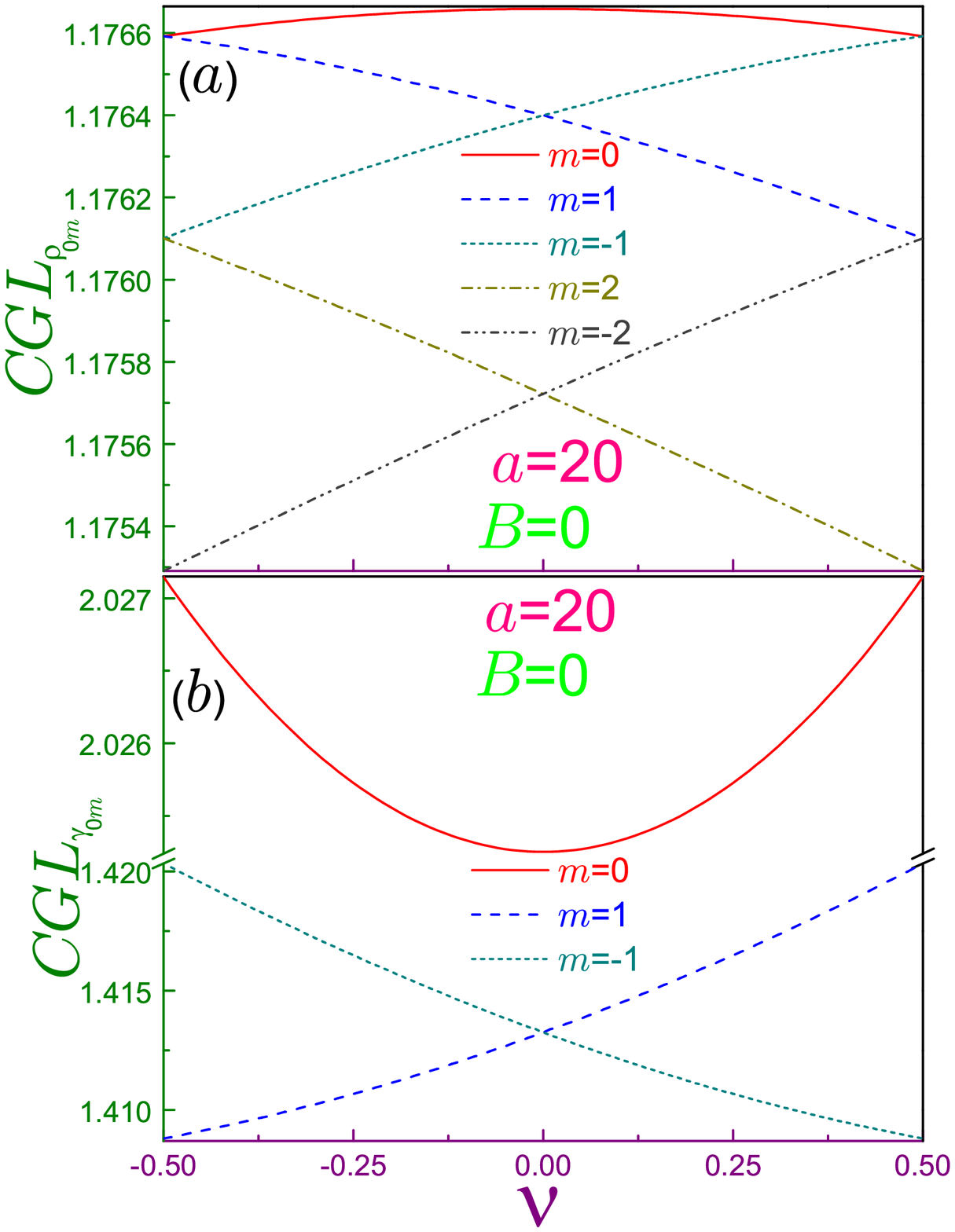}
\caption{\label{CGLabFig1}(a) Position ${CGL_\rho}_{0m}$ and (b) momentum ${CGL_\gamma}_{0m}$ complexities as functions of the normalized AB flux $\nu$ for several magnetic quantum numbers $m$ denoted in each of the corresponding subplot. The same parameters as in Fig.~\ref{ShannonFig3} are used. Note vertical axis break in panel (b) from 1.42033 to 2.02524.}
\end{figure}

Dependence of the complexities ${CGL_\rho}_{0m}$ and ${CGL_\gamma}_{0m}$ on the AB flux is illustrated in panels (a) and (b) of Fig.~\ref{CGLabFig1}, respectively. It is seen that qualitatively they reproduce the Onicescu energies behaviour. As it follows from Eqs.~\eqref{Shannon6} and \eqref{Onicescu6}, the ground-state position part for the thin rings is a concave function of $\phi_{AB}$:
\begin{equation}\label{CGL2}
{CGL_\rho}_{00}(\nu)=\left(\frac{e}{2}\right)^{1/2}\left(1-\frac{\nu^2}{48a^{3/2}}\right),\quad a\gg1.
\end{equation}
For the larger ring radius, the variation of the complexities with the flux $\phi_{AB}$, similar to all other measures, is being suppressed; for example, at $a\rightarrow\infty$ the lowest level position component approaches, as it directly follows from Eq.~\eqref{CGL2}, the value of $(e/2)^{1/2}$, irrespectively of the AB intensity.

\begin{figure}
\centering
\includegraphics[width=\columnwidth]{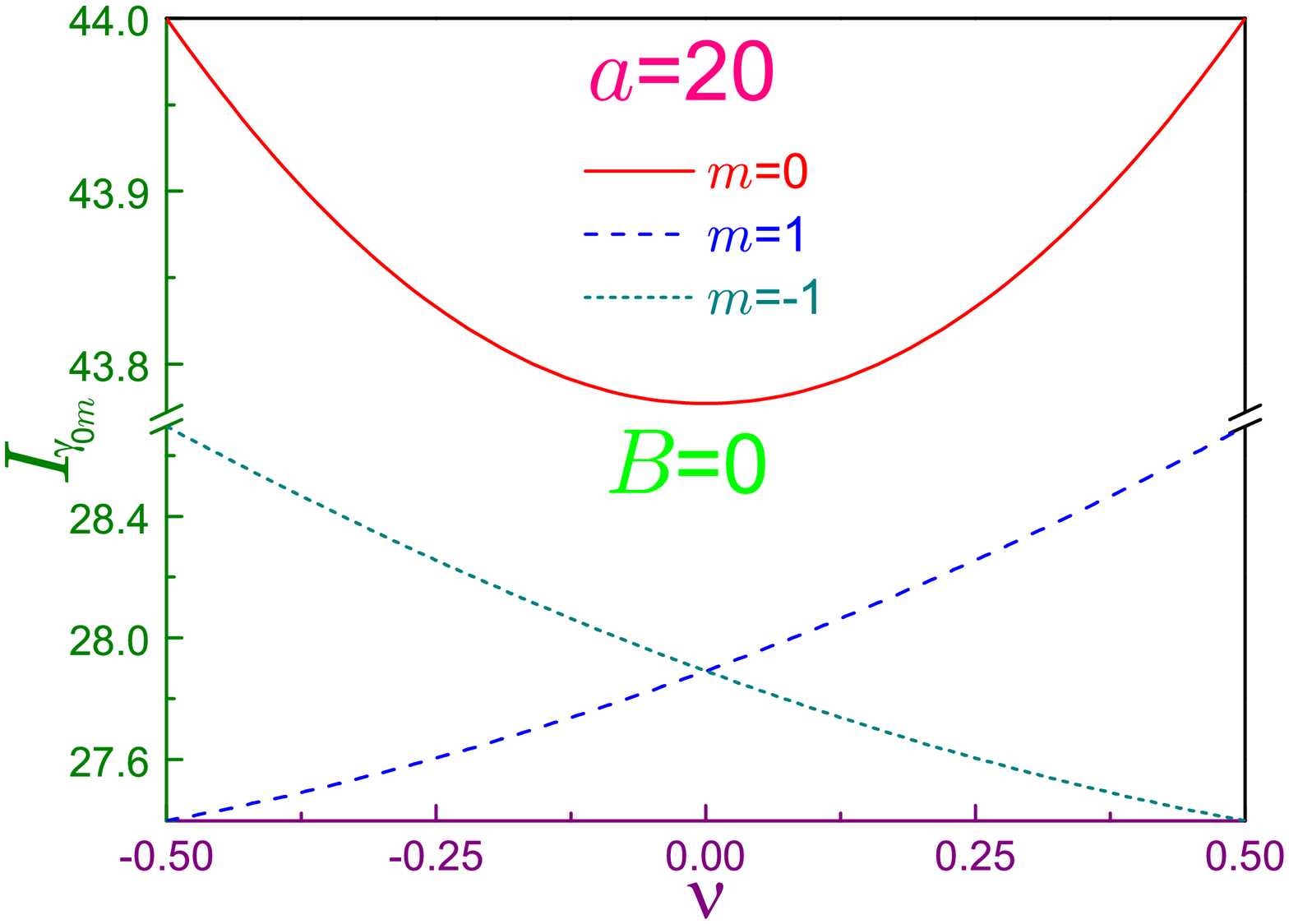}
\caption{\label{FisherFig1}Momentum Fisher informations ${I_\gamma}_{0m}$ as functions of the normalized AB flux $\nu$ for several azimuthal numbers $m$ where solid curve corresponds to $m=0$, and dashed (dotted) line is for the $m=1$ ($m=-1$) state. The same parameters as in Fig.~\ref{ShannonFig3} are used. Note vertical axis break from 28.696 to 43.772.}
\end{figure}

Next, as Fig.~\ref{FisherFig1} demonstrates, momentum Fisher information ${I_\gamma}_{00}$ has its minimum at the zero AB flux and, similar to all other $m=0$ measures is symmetric with respect to $\nu=0$ whereas for the orbitals with the positive (negative) magnetic index $m$ it increases (decreases) with $\nu$. In addition, as mentioned above, the degeneracy expressed by equations~\eqref{Degeneracy1} for the energies, does not hold for the momentum component ${I_\gamma}_{nm}$. Since the position components depend neither on $B$ nor $\phi_{AB}$, all these properties are also characteristic of the product ${I_\rho}_{nm}{I_\gamma}_{nm}$.

\begin{figure}
\centering
\includegraphics[width=\columnwidth]{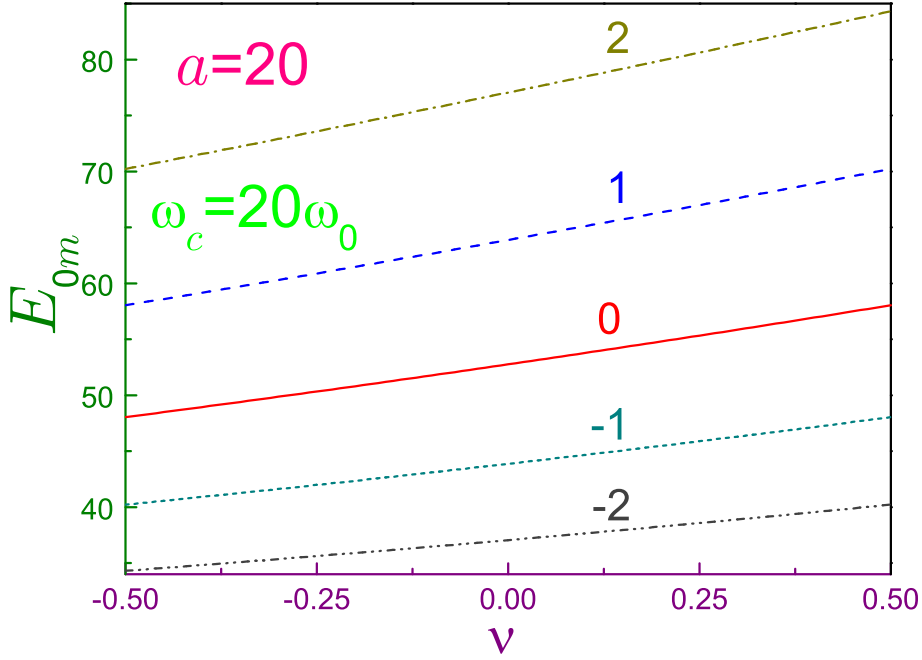}
\caption{\label{EnergiesFig1}Energy spectrum $E_{0m}$ (in units of $\hbar\omega_0$) as a function  of the normalized AB flux $\nu$ for several azimuthal indices $m$ denoted by the number near the corresponding curve. QAD strength is $a=20$ and ratio $\omega_c/\omega_0=20$.}
\end{figure}

As a last remark, let us point out that in the presence of the uniform magnetic intensity, $B\neq0$, the quantum-information measures preserve the shape of their $B=0$ dependence on the AB flux, as is seen, for example, from Eqs.~\eqref{Shannon5},  \eqref{Onicescu5} and \eqref{CGL2} for the Shannon entropy, Onicescu energy and complexity. However, the structure of the energy spectrum is drastically modified by the field $B$; namely, since it introduces the preferred azimuthal direction, the energies $E_{n0}$ are not anymore symmetric functions of $\phi_{AB}$ demonstrating instead at the quite strong $B$ a continuous increase with the dimensionless flux $\nu$. The same is correct for the orbitals with the nonzero magnetic index $m$ too. A representative example of the field such that $\omega_c/\omega_0=20$ is shown in Fig.~\ref{EnergiesFig1}. The invariance under the transformation from Eq.~\eqref{Transformation1} is not destroyed by the uniform field but the degeneracies from Eqs.~\eqref{Degeneracy1} are lifted by it.

\section{Conclusions}
Analysis of the quantum-information measures of the nanostructure requires a correct evaluation of the corresponding one-particle waveforms. Their position components $\Psi_{nm}({\bf r})$ for the 2D electron in the QR subject to the superposition of the transverse uniform magnetic field and the AB flux have been known very well \cite{Bogachek1,Tan1,Tan2,Tan3,Fukuyama1,Olendski3}. Above, momentum wave functions $\Phi_{nm}({\bf k})$ have been calculated straightforwardly, Eq.~\eqref{MomentumFUnction1}, as Fourier transforms of their position counterparts, Eq.~\eqref{Fourier1_1}. It was shown, in particular, that the increasing intensity $B$ subdues oscillations of the radial part of $\Phi_{nm}({\bf k})$ transforming them in the limit of the infinitely strong fields into the flat zero-amplitude surface, Eq.~\eqref{FunctionLimits1_1}. Importantly, in the case of the AB-free QD the expression for the ground-band momentum waveform is nothing else but the modified Gaussian, Eq.~\eqref{MomentumFunction3_1}, what means, that its lowest orbital, $m=0$, does saturate Shannon entropy uncertainty relation, Eq.~\eqref{EntropyInequality1}. Obviously, all other states satisfy this relation too as an inequality in which the sum of the two Shannon entropies ${S_\rho}_{nm}+{S_\gamma}_{nm}$ is a $B$-independent quantity since the first and second items in it contain $\pm\ln\left(1+\frac{1}{4}\frac{\omega_c^2}{\omega_0^2}\right)$, respectively. Other main findings include independencies of the position Fisher information on the azimuthal quantum number $m$ and of the products ${I_\rho}_{nm}{I_\gamma}_{nm}$ and ${O_\rho}_{nm}{O_\gamma}_{nm}$ and both position ${CGL_\rho}_{nm}$ and momentum ${CGL_\gamma}_{nm}$ parts of the complexity on the uniform field $B$. A comparison of the analytic expressions at $B=0$ of the dependencies of the ground-state position Shannon entropy, Onicescu energy and  the energy spectrum on the AB flux reveals a strong similarity of the former and latter variations what can be used in deducing the magnitude of either of them if the other quantity is known; in particular, from the measurement of the persistent current one can infer what Shannon entropy and Onicescu energies are.

In addition to the measures discussed above, in a similar way the quantities
\begin{subequations}\label{Uncertainty1}
\begin{align}\label{Uncertainty1_R}
\left\langle r^2\right\rangle&=\int_{\mathbb{R}^l}r^2\rho({\bf r})d{\bf r}\\
\label{Uncertainty1_K}
\left\langle k^2\right\rangle&=\int_{\mathbb{R}^l}k^2\gamma({\bf k})d{\bf k}
\end{align}
\end{subequations}
that enter the uncertainty relation \cite{Bracher1}
\begin{equation}\label{UncertaintyRelation1}
\sqrt{\left\langle r^2\right\rangle}\sqrt{\left\langle k^2\right\rangle}\geq|m|+1
\end{equation}
can be considered too; in particular, explicit evaluation yields for the position component:
\begin{equation}\label{Uncertainty2R}
\left\langle r^2\right\rangle=2(2n+\lambda+1)r_{eff}^2.
\end{equation}
Its momentum counterpart
\begin{subequations}\label{Uncertainty2K}
\begin{align}\label{Uncertainty2K_1}
&\left\langle k^2\right\rangle=\frac{1}{r_{eff}^2}\frac{n!}{\Gamma(n\!+\!\lambda\!+\!1)}\int_0^\infty\!\!\!d\xi\xi^3\!\left[\int_0^\infty\!\!\!\!e^{-z/2}z^{\lambda/2}L_n^\lambda(z)J_{|m|}\!\left(\!2^{1/2}\xi z^{1/2}\right)\!dz\right]^2
\intertext{can be derived analytically for the lowest band, $n=0$, of the QD ($a=0$) without the AB flux:}
\label{Uncertainty2K_2}
&\left.\left\langle k^2\right\rangle_{0m}\right|_{a=\nu=0}=\frac{1}{r_{eff}^2}\frac{|m|+1}{2}.
\end{align}
\end{subequations}
From Eqs.~\eqref{Uncertainty2R} and \eqref{Uncertainty2K_1} it is seen that the product $\left\langle r^2\right\rangle\left\langle k^2\right\rangle$ is a $B$-independent quantity again. Let us also note that the lowest band of the AB-free QD does saturate the uncetainty relation, Eq.~\eqref{UncertaintyRelation1}, since the position and momentum waveforms represent in this case modified Gaussian functions \cite{Bracher1}, see, e.g., Eq.~\eqref{MomentumFunction3_1}.

Finally, let us point out that the wave functions introduced here can be used in the analysis of the R\'{e}nyi \cite{Renyi1,Renyi2} $R_{\rho,\gamma}$ and Tsallis \cite{Tsallis1} $T_{\rho,\gamma}$ entropies whose expressions in terms of the densities $\rho({\bf r})$ and $\gamma({\bf k})$ are:
\begin{subequations}
\label{Functionals1}
\begin{align}\label{RenyiX1}
R_\rho(\alpha)&=\frac{1}{1-\alpha}\ln\!\left(\int_{\mathbb{R}^l}\rho^\alpha({\bf r})d{\bf r}\right)\\
\label{RenyiP1}
R_\gamma(\alpha)&=\frac{1}{1-\alpha}\ln\!\left(\!\int_{\mathbb{R}^l}\gamma^\alpha({\bf k})d{\bf k}\right)\\
\label{TsallisX1}
T_\rho(\alpha)&=\frac{1}{\alpha-1}\left(1-\int_{\mathbb{R}^l}\rho^\alpha({\bf r})d{\bf r}\right)\\
\label{TsallisP1}
T_\gamma(\alpha)&=\frac{1}{\alpha-1}\left(1-\int_{\mathbb{R}^l}\gamma^\alpha({\bf k})d{\bf k}\right).
\end{align}
\end{subequations} 
For each bound quantum orbital, the position and momentum components of these one-parameter functionals are not independent from each other but obey the following uncertainties \cite{Bialynicki2,Zozor1,Rajagopal1}:
\begin{subequations}\label{RenyiTsallisUncertainty1}
\begin{align}\label{RenyiUncertainty1}
&R_\rho(\alpha)+R_\gamma(\beta)\geq-\frac{l}{2}\left(\frac{1}{1-\alpha}\ln\frac{\alpha}{\pi}+\frac{1}{1-\beta}\ln\frac{\beta}{\pi}\right)\\
\label{TsallisInequality1}
&\left(\frac{\alpha}{\pi}\right)^{l/(4\alpha)}\!\!\left[1+(1-\alpha)T_\rho(\alpha)\right]^{1/(2\alpha)}\geq\left(\frac{\beta}{\pi}\right)^{l/(4\beta)}\!\!\left[1+(1-\beta)T_\gamma(\beta)\right]^{1/(2\beta)},
\end{align}
\end{subequations}
where the non-negative coefficients $\alpha$ and $\beta$ obey the constraint:
\begin{equation}\label{RenyiUncertainty2}
\frac{1}{\alpha}+\frac{1}{\beta}=2,
\end{equation}
which for the case of the Tsallis relation, Eq.~\eqref{TsallisInequality1}, has to be supplemented by the additional requirement:
\begin{equation}\label{Sobolev2}
\frac{1}{2}\leq\alpha\leq1.
\end{equation}
In the limit $\alpha\rightarrow1$ (and, accordingly, $\beta\rightarrow1$) Eqs.~\eqref{Functionals1} and \eqref{RenyiUncertainty1} degenerate into their Shannon counterparts, Eqs.~\eqref{Shannon1} and \eqref{EntropyInequality1}, respectively, whereas the Tsallis inequality turns into the identity with each of its sides equal to $\pi^{-l/4}$. Recently, it was conjectured \cite{Olendski4} that the lowest orbital of the $l$-dimensional quantum structure in the limit $\alpha=1/2$ transforms R\'{e}nyi, Eq.~\eqref{RenyiUncertainty1}, and Tsallis, Eq.~\eqref{TsallisInequality1},  uncertainty relations into the equalities. One can check this conjecture for the AB-free ($\nu=0$) QD ($a=0$) lowest band ($n=0$) for which, for example, R\'{e}nyi entropies read:
\begin{subequations}\label{Renyi2}
\begin{align}\label{Renyi2_X}
\left.{R_\rho}_{0m}(\alpha)\right|_{a=\nu=0}&=2\ln r_{eff}+\ln2\pi+\frac{1}{1-\alpha}\ln\frac{\Gamma(|m|\alpha+1)}{(|m|!)^\alpha\alpha^{|m|\alpha+1}}\\
\label{Renyi2_K}
\left.{R_\gamma}_{0m}(\alpha)\right|_{a=\nu=0}&=-2\ln r_{eff}+\ln\frac{\pi}{2}+\frac{1}{1-\alpha}\ln\frac{\Gamma(|m|\alpha+1)}{(|m|!)^\alpha\alpha^{|m|\alpha+1}}.
\end{align}
\end{subequations}
Asymptote $\alpha\rightarrow1/2$ degenerates right-hand side of Eq.~\eqref{RenyiUncertainty1} into $2\ln2\pi$ whereas its left-hand side for the functionals from Eqs.~\eqref{Renyi2} yields in the same limit:
\begin{equation}\label{Renyi3}
\left.[{R_\rho}_{0m}(\alpha)+{R_\gamma}_{0m}(\beta)]\right|_{a=\nu=0}=2\ln2\pi+|m|(1+\ln2)+\ln\!\frac{\Gamma^2\!\left(\!\frac{|m|}{2}+1\right)}{|m|^{|m|}}.
\end{equation}
As a function of the magnetic quantum number, the expression from Eq.~\eqref{Renyi3} monotonically increases with $|m|$ and for the azumthally symmetric state, $m=0$, which is the lowest-energy orbital, it converts R\'{e}nyi inequality, Eq.~\eqref{RenyiUncertainty1}, into the identity confirming in this way the above mentioned conjecture. For the QR, $a\neq0$, an analysis of the R\'{e}nyi and Tsallis entropies requires separate careful consideration.

\section{Acknowledgments}
Research was supported by SEED Project No. 1702143045-P from the Research Funding Department, Vice Chancellor for Research and Graduate Studies, University of Sharjah.

\bibliographystyle{model1a-num-names}

\end{document}